\documentclass[]{interact}
\usepackage{epstopdf}
 \usepackage[caption=false,position=top]{subfig}

\usepackage[numbers,sort&compress]{natbib}
\usepackage{url}

\usepackage{comment}

\theoremstyle{plain}

\theoremstyle{definition}

\theoremstyle{remark}

\begin{document}

\title{Trends and gender disparities in grades and grade penalties among bioscience and health-related major students before, during, and after COVID-19 remote instruction}

\author{
  Alysa Malespina,
  Fargol Seifollahi$^{*}$\thanks{$^{*}$Corresponding author. Email: fas102@pitt.edu},
  Chandralekha Singh\\
  \textit{Department of Physics \& Astronomy, University of Pittsburgh, Pittsburgh, PA, USA 15260}
}

\maketitle

\begin{abstract}

In this study, we investigate student performance using grades and ``grade anomalies'' across periods before, during, and after COVID-19 remote instruction in courses for bioscience and health-related majors. Additionally, we explore gender equity in these courses using these measures. We define grade anomaly as the difference between a student's grade in a course of interest and their overall grade point average (GPA) across all other courses taken up to that point. If a student's grade in a course is lower than their GPA in all other courses, we refer to this as a ``grade penalty''. Students received grade penalties in all courses studied, consisting of twelve courses taken by the majority of bioscience and health-related majors.  Overall, we found that both grades and grade penalties improved during remote instruction but deteriorated after remote instruction. Additionally, we find more pronounced gender differences in grade anomalies than in grades.   We hypothesize that women’s decisions to pursue STEM careers may be more influenced by the grade penalties they receive in required science courses than men’s, as women tend to experience larger penalties across all periods studied. Furthermore, institutions concerned with equity should consider grade penalties as a straightforward measure and make a conscious effort to consider their implications.

\end{abstract}


\section{Introduction}

In the wake of the COVID-19 pandemic, many education researchers have been focusing on assessing differences between online and in-person courses regarding student learning outcomes and classroom equity \cite{marzoli2022, palmgren2022, kortemeyer2022, klumpp2021}.  There are mixed findings regarding the effect of online instruction on student learning \cite{marzoli2022, palmgren2022, gonzalez2020}. In this study, we explore overall trends in both grades and grade anomalies before, during, and after remote instruction due to COVID-19 in courses for bioscience and health-related majors. We define a grade anomaly as the difference between a student’s grade in a specific course and their grade point average (GPA) in all other classes up to that point. The mean of this metric across all students who took a course is referred to as the average grade anomaly (AGA).  We categorize AGAs into ``bonuses'' and ``penalties''.  A course in which students on average earn a lower grade than usual has an AGA with a grade penalty, while a course in which students on average earn a higher grade than usual has an AGA with a grade bonus.

Within our framework, we suggest that grade anomalies may serve as an important indicator of how courses influence students' academic self-concept, as observed through institutional grade data. Academic self-concept is a relatively stable measure of a student's perceived ability to succeed in the academic sphere, and is based on grades and outside feedback (e.g., from parents, peers, and instructors) \cite{spence1983, gniewosz2012, wigfield2002, eccles2020, wigfield2010}.  Grades inform academic self-concept as both an external (``How good at math am I compared to other students?'') and internal (``How good am I at math compared to English?'') frame of reference \cite{spence1983, gniewosz2012, wigfield2002}.  We also note that, while academic self-concept is generally quite stable, it is possible for it to change quickly during periods of transition (such as the transition from high school to university) \cite{gniewosz2012}. It has been shown that grade penalties in STEM courses during the first two semesters (but not later semesters) of university were negatively correlated with  completing a STEM degree, even when controlling for gender, race, high school preparation, and college performance~\cite{witteveen2020}. These findings hint at the importance of monitoring and minimizing grade penalties in students' first few semesters.

\section{Literature Review}

\subsection{Grade Anomaly Concept and the Rationale}

Our framework uses the grade anomaly as a central construct instead of the grade because students’ academic self-concept is typically shaped by relative comparisons rather than absolute grades \cite{eccles2020}. Students may compare their grades across courses to determine which disciplines they excel at or struggle with \cite{eccles2020}. Additionally, students tend to have a fairly fixed view of what “kind” of student they are; e.g., students may endorse the idea that ``If I get As, I must be an A kind of person. If I get a C, I am a C kind of person” \cite{seymour2019}. Grade anomalies may challenge or reinforce students’ ideas about what kind of student they are, and if they are capable of succeeding in their chosen major. Many students who leave STEM majors explicitly cite lower grades than they are used to as a reason for doing so \cite{seymour1997,seymour2019}.  Grade penalties are more common and extreme in STEM disciplines than in humanities or social science departments \cite{seymour2019, rask2010, koester2016, matz2017}.

\subsection{COVID-19 Pandemic and Student Learning}

There have been mixed findings about the impact of the COVID-19 pandemic on student learning and academic performance. A study conducted in K-12 schools using web surveys to explore school-related experiences among middle and high school students revealed that academic grades were negatively impacted during the COVID-19 pandemic~\cite{fisher2022student}. While this study showed that the students participating in remote or hybrid learning were more likely to experience academic decline compared to their peers with in-person instruction \cite{fisher2022student}, another study focused on high school students in online chemistry classes reported a statistically significant improvement in students' critical thinking skills and overall academic achievement during the COVID-19 online classes period \cite{almahdawi2021high}. At the college level, students' academic performance in a virtual General Chemistry 2 course showed a decline compared to previous in-person class averages, which was likely attributed to factors such as difficulty adapting to online learning, diminished peer interactions, and reduced utilization of office hours due to the COVID-19 pandemic \cite{goyal2022student}. Prior studies investigating college physics students' responses to remote learning have shown a decreased engagement and peer interaction, as well as lower reported self-organization skills, i.e., the extent to which students take a proactive role in monitoring their own learning \cite{marzoli2022, klein2021studying}. A comparative study of science and engineering students' performance in an introductory physics class before and after COVID-19 showed only a few significant differences between students' pre- and post-pandemic learning behavior and academic achievement, with the significant differences being small effects \cite{nemeth2023comparing}. Additionally, an investigation into grades and grade anomalies for required courses for engineering students at a large urban institution revealed an overall positive trend during the remote instruction period \cite{malespina2023_eng_anomalies_covid}.

\subsection{Gender Disparities in STEM Education and Career Outcomes}

Previous research has explored gender disparities in STEM subjects, highlighting differences in how these disparities are perceived by instructors and students \cite{makarova2019, musters2024views}. Additionally, gender differences in performance and persistence in science, technology, engineering, and mathematics (STEM) have been closely studied in fields such as physics or engineering, in which women are under-represented \cite{gonzalez2012, goodman2002, whitcomb2020a, whitcomb2021b, sawtelle2012, cavallo2004, marshman2018, nissen2016, pintrich1990}. This line of research has been less common in fields such as biology, in which women are not under-represented \cite{ballen2017b,haak2011}.  However, even in fields in which  women and men earn similar numbers of undergraduate degrees \cite{nces2020}, they may have very different classroom experiences \cite{good2012, raelin2014, eaton2020, cwik2022}.   

There are many examples of gender inequities in biological science classrooms as well as career paths. Women in biology classrooms are less likely to participate in classroom discussions, are viewed as less knowledgeable by their peers, and tend to have lower exam grades in introductory biology courses \cite{eddy2014, grunspan2016, ballen2017a}.  After graduation, women with Biological Science graduate degrees are less likely than men to work as scientists \cite{elliott2016}. If they pursue jobs in research, women in biology tend to have shorter publishing careers and lower yearly publishing rates than men in the same field~\cite{huang2020, holman2018}. If they choose to pursue medical careers, women may still experience inequities: there are gender disparities in compensation and time to promotion for all academic medical specialties \cite{aamc2021faculty} as well as for physicians \cite{ly2016}. If gender differences in career outcomes are not explained by a lack of representation in the classroom, we hypothesize that academic self-concept may provide some insight \cite{spence1983, gniewosz2012, wigfield2002}. Low academic self-concept may lead to lower future achievement and persistence because it discourages student engagement in a domain \cite{spence1983}. When women leave STEM disciplines, they often do so with higher grades than the men who remain in the program \cite{maries2022, seymour1997, seymour2019}. 

The overall trends in grades and grade anomalies for students in physics, engineering, and bioscience majors have been examined \cite{malespina2022gender, malespina2022impact,malespina2023_bio_grade_anomalies}. Additionally, researchers have explored how these trends and gender differences varied before, during, and after remote instruction due to the COVID-19 pandemic specifically for engineering majors, where men are the majority \cite{malespina2023_eng_anomalies_covid}. These studies have shown that women often get larger grade penalties than men for courses required for their major. However, there has been limited focus on the trends before, during, and after COVID-19 for students in bioscience and health-related majors, in which women outnumber men.

Broadly, in this research we aim to understand differences in grade anomalies before, during, and after remote instruction due to COVID-19 for bioscience major students with a particular focus on gender differences in grades and grade anomalies. This study builds on previous work which observed grade anomalies at this same institution for over ten years pre-COVID \cite{malespina2023_bio_grade_anomalies, malespina2023investigating} with a similar methodology. We aim to answer the following research questions regarding grade anomalies: 

\begin{enumerate}
    \item[RQ1.] Do grades or grade anomalies differ before, during, and after remote instruction due to the COVID-19 pandemic?
    \item[RQ2.]Are there gender differences in grades or grade anomalies, and do they differ before, during, and after remote instruction due to the COVID-19 pandemic? 
\end{enumerate}

\section{Methodology}

\subsection{Participants}

Participants in this study were enrolled in bioscience or health majors at a large, public, urban institution. The student major breakdown can be found in Table \ref{majors}. All majors except for Neuroscience, Pharmacy, and Rehabilitation Science are offered through the Department of Biological Sciences. These students were chosen because of their similar course requirements, especially for large introductory science courses. 

Grade data were collected over four years. We divide these semesters into three groups, which are described in Table \ref{periods}. This study uses cross-sectional data, meaning that we are not tracking the same group of students over time and the samples are independent. We excluded courses that were taken during the summer semester. We excluded summer courses because they are not a typical representation of courses at our institution. For instance, many students enrolled in the summer courses are local individuals who are visiting home for the summer and do not primarily attend this institution. Additionally, class sizes during the summer are an order of magnitude smaller compared to those in the Fall and Spring semesters.

\begin{table}[tb]
\tbl{Major information for study participants \label{majors}}
{\begin{tabular} {l|ccc} \toprule
Major & Pre-Remote & Remote & Post-Remote \\ \midrule
 Biological Sciences &  29\% & 24\% & 21\% \\
 Bioinformatics/Computational Biology &  2\% & 2\% & 1\% \\
 Ecology and Evolution &  1\% & 1\% & 1\% \\
 Microbiology &  5\% & 5\% & 3\% \\
 Molecular Biology&  5\% & 4\% & 4\% \\
 Neuroscience &  22\% & 17\% & 12\% \\
 Pharmacy&  1\% & 0\% & 0\% \\
 Rehabilitation Science &  7\% & 5\% & 3\% \\ 
 Undeclared &  28\% & 42\% & 55\% \\ \bottomrule
\end{tabular}}
\raggedright 
\bigskip 
\small\textit{Note.} Students who have not chosen a major are included because students are not required to declare a major until their third year.  Undeclared students who later went on to declare any major not on this list were excluded. The percent of undeclared majors increases over time because students in later periods of the study have not been enrolled as long as students in earlier periods. 
\end{table}

\begin{table}[tb]
\tbl{Labels for each time period studied \label{periods}}
{\begin{tabular} {l|lll} \toprule
Label & \multicolumn{1}{c}{Period} \\ \midrule
Pre-Remote & Four semesters of in-person instruction before the COVID-19 pandemic, excluding Spring 2020  \\
Remote & Two semesters of remote instruction due to the COVID-19 pandemic, excluding Spring 2020 \\
Post-Remote & Two semesters of in-person instruction after the return to in-person classes \\
 \bottomrule
\end{tabular}}
\end{table}

This left us with 14,149 pre-remote, 6145 remote, and 5484 post-remote instances of enrollment in a course. For example, a student who takes four courses in one semester and three in the next semester has seven instances of enrollment. Demographic information for the student sample can be found in Table \ref{demographics}. This research was carried out in accordance with the principles outlined in the Institutional Review Board (IRB) ethical policy of the institution, and de-identified demographic data were provided through university records.

\begin{table}[tb]
\tbl{Demographic information for study participants \label{demographics}}
{\begin{tabular} {l|cc|cccccc} \toprule
& \multicolumn{2}{c}{Sex} & \multicolumn{6}{c}{Race/Ethnicity} \\
Period & Female & Male & Asian & Black & Latine & Multiracial & White & Unknown \\ \midrule
Pre-Remote & 61\% & 39\% & 25\% & 5\% & 5\% & 5\% & 59\% &  1\% \\ 
Remote & 63\% & 37\% & 27\% & 5\% & 6\% & 5\% & 56\% & 1\% \\
Post-Remote & 60\% & 40\% & 26\% & 5\% & 6\% & 5\% & 57\%  & 1\%\\
\bottomrule
\end{tabular}}
\raggedright
\bigskip 
\small\textit{Note.} Two ethnicity groups, Indigenous American and Pacific Islander, were excluded from this table because they each made up less than 0.5\% of the sample.
\end{table}

\subsection{Course Selection}

We chose to study courses that were taken by the largest number of students, excluding non-major electives (for example, ``Introduction to Piano'' or ``Public Speaking'') and courses that make up general education requirements. Thus, many courses were mandatory for students in the majors on which we focus. However, not all courses were required for students in all the majors in our sample. Information about whether a course was required, optional (i.e., an elective that counts towards the major), or not required is included in Table \ref{course_requirements} in the appendix. The courses we chose are listed in Table \ref{semester} in the appendix, along with information about the year in which the students typically take the course. We also note that, although Human Physiology is not required for most majors studied, it met our criteria because it is a commonly chosen elective for Biological Science students.

\subsection{Measures}
\subsubsection{Course Grade} Course grades were based on a 0--4 scale used at this university, and the conversion of letter grades to GPA points can be seen in Table \ref{grade_conversion}.  We are unable to report grading schemes of each instructor, type of course (i.e., traditional lectures or active learning), or any other detailed course-level information due to the large number of courses sampled.

\begin{table}[tb]
\tbl{Grades and GPA points for this university's grading standards. For all majors included here at this institution, a C or above is a passing grade. \label{grade_conversion}}
{\begin{tabular} {l|cccccccccccc} \toprule
Grade & A/A$+$ & A$-$ & B$+$ & B & B$-$ & C$+$ & C & C$-$ & D$+$ & D & D$-$ & F \\ \midrule
GPA Value & 4.00 & 3.75 & 3.25 & 3.00 & 2.75 & 2.25 & 2.00 & 1.75 & 1.25  & 1.00 & 0.75 & 0.00\\ 
 \bottomrule
    \end{tabular}}
\raggedright
\bigskip 
\end{table}

\subsubsection{Grade Anomaly} Grade anomaly (GA) was found by first finding each student’s grade point average excluding the course of interest ($GPA_{exc}$). This was performed by using the equation
\begin{equation}
    GPA_{exc} = \frac{ GPA_{c} \times Units_{c} - Grade \times Units}{Units_{c} - Units}
\end{equation}
where $GPA_{c}$ is the student's cumulative GPA, $Units_{c}$ is the cumulative number of units the student has taken, $Grade$ is the grade the student received in an individual course, and $Units$ (also called credit hours) is the number of units associated with an individual course.  After finding $GPA_{exc}$, we can calculate grade anomaly by finding the difference between a student's $GPA_{exc}$ and the grade received in that class:
\begin{equation}
    GA = Grade - GPA_{exc}.
\end{equation}

A negative GA corresponds to a course grade lower than a student's GPA in other classes and we call this a ``grade penalty''. A positive GA corresponds to a course grade higher than a student's GPA in other classes and we call this a ``grade bonus''. Average grade anomaly (AGA) is the mean of students’ grade anomalies for each course, and is the metric by which we compare courses.

\subsubsection{Analysis}

The analysis was conducted using university data that was made available to us. We report sample size, mean, and standard deviation of each measurement for each course of interest in order to characterize both average grade anomaly (AGA) and grades. These statistics were computed separately for women and men, as well as for all students combined.  We also compared the effect size of gender on both grade and grade anomaly, using Cohen's d to describe the size of the mean differences and unpaired \textit{t}-tests to evaluate the statistical robustness of the differences.  Cohen's \textit{d} is calculated as follows: 
\begin{equation}
    d = \frac{\mu_{1} - \mu_{2}}{\sqrt{{(\sigma_{1}^{2}+\sigma_{2}^{2})}/2}}
\end{equation}
where $\mu_{1}$ and $\mu_{2}$ are the means of the two groups, $\sigma_{1}$ and $\sigma_{2}$ are the standard deviations~\cite{frey2018}, and Cohen's \textit{d} is considered small if $d$$\sim$0.2, medium if $d$$\sim$0.5, and large if $d$$\sim$0.8~\cite{cohen1988}. We used a significance level of 0.05 in the independent \textit{t}-tests as a balance between Type I (falsely rejecting a null hypothesis) and Type II (falsely accepting a null hypothesis) errors  \cite{frey2018}. Given the distributional characteristics of our data with slight skewness yet large sample size, the t-test results remain robust to the violations of normality assumption. 
Analysis was conducted using R \cite{r}, using the package plotrix \cite{plotrix} for descriptive statistics, lsr \cite{lsr} for effect sizes, and ggplot2 \cite{ggplot2} to create plots.

\section{Results}

\begin{figure}
\centering
\subfloat[ \label{fig:grade_combo_precovid}]{%
{\includegraphics[width=1.0\textwidth]{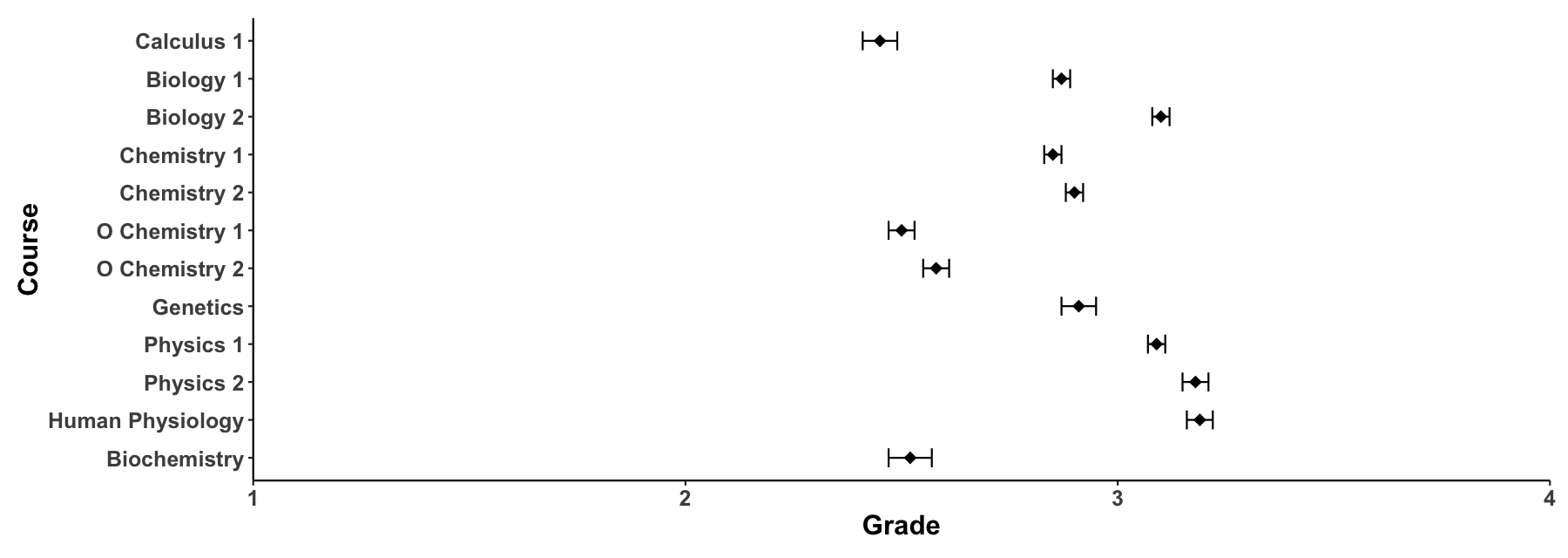}}}\vspace{30pt}
\subfloat[ \label{fig:grade_combo_remote}]{%
{\includegraphics[width=1.0\textwidth]{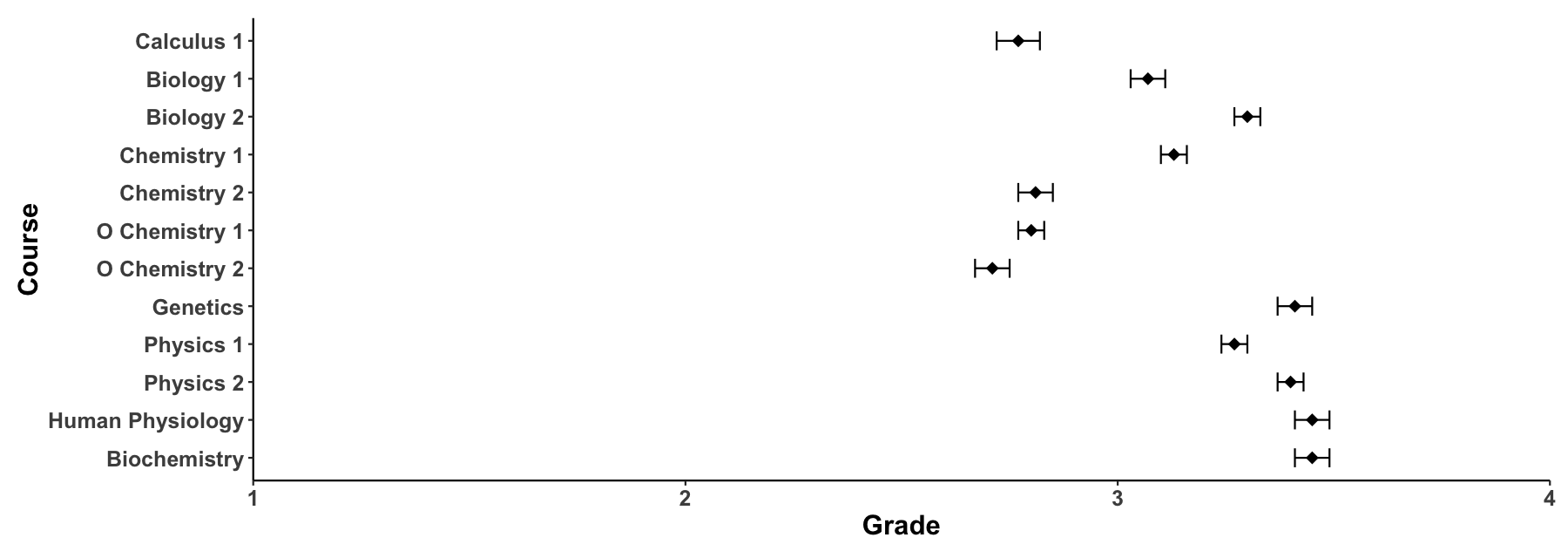}}}\vspace{30pt}
\subfloat[ \label{fig:grade_combo_inperson}]{%
{\includegraphics[width=1.0\textwidth]{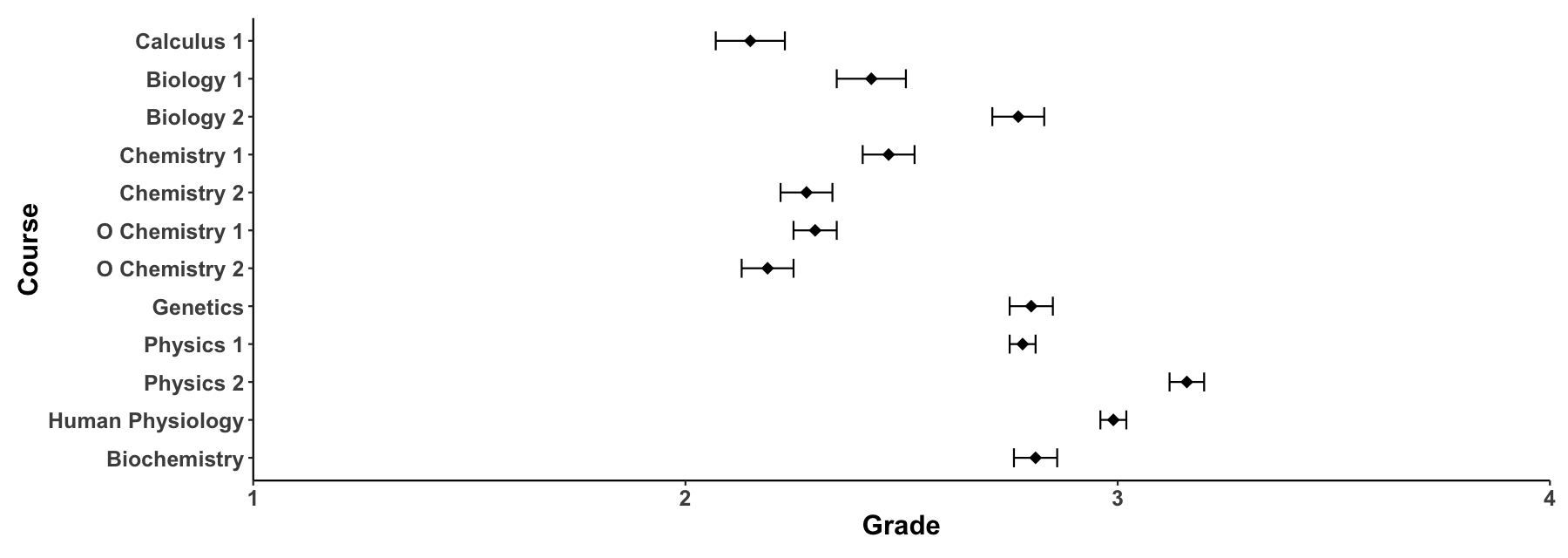}}}
\caption{Comparison of student grades for each course of interest for classes before (a), during (b), and after (c) remote instruction due to the COVID-19 pandemic. Ranges represent standard error of the mean.} \label{fig:grade_combo}
\end{figure}

The primary result from our study in regard to research question 1 is the following: course grades were generally the highest during remote instruction compared to grades before or after, which can be seen in Figure \ref{fig:grade_combo}. Additionally, post-remote grades tended to be either the same or a fraction of a letter grade lower than pre-remote grades, which can be seen in either Figure \ref{fig:grade_combo} or in Table \ref{grade_combo_table} in the appendix. For example, if a pre-remote instruction average grade was a C$+$, it is likely to be a C$+$ or a C after remote instruction.  Also regarding research question 1, AGAs had the smallest magnitude during remote classes, which can be seen in Figure \ref{fig:aga_combo} or Table \ref{AGA_combo_table} in the appendix, and most courses had larger magnitudes of AGAs post-remote instruction compared to pre-remote instruction. Given that all AGAs are negative, by the terms smaller or larger, we refer to the magnitude throughout this paper.

\begin{figure}
\centering
\subfloat[\label{fig:grade_gender_precovid}]{%
{\includegraphics[width=1.0\textwidth]{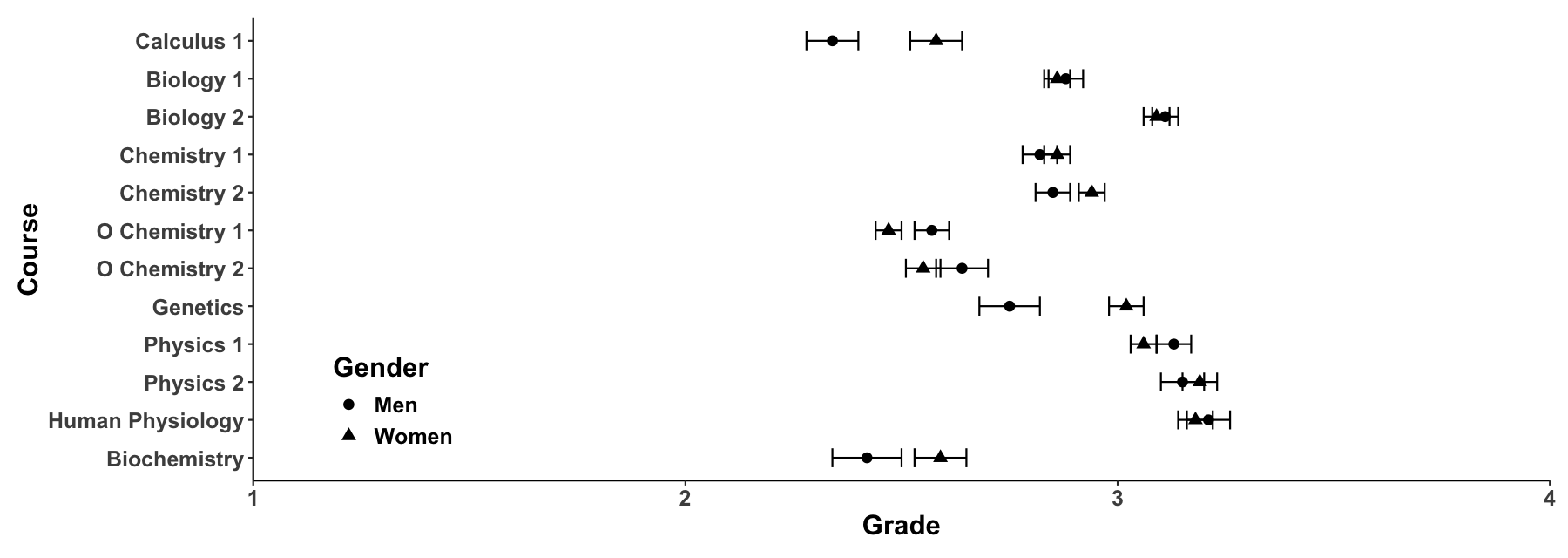}}}\vspace{30pt}
\subfloat[ \label{fig:grade_gender_remote}]{%
{\includegraphics[width=1.0\textwidth]{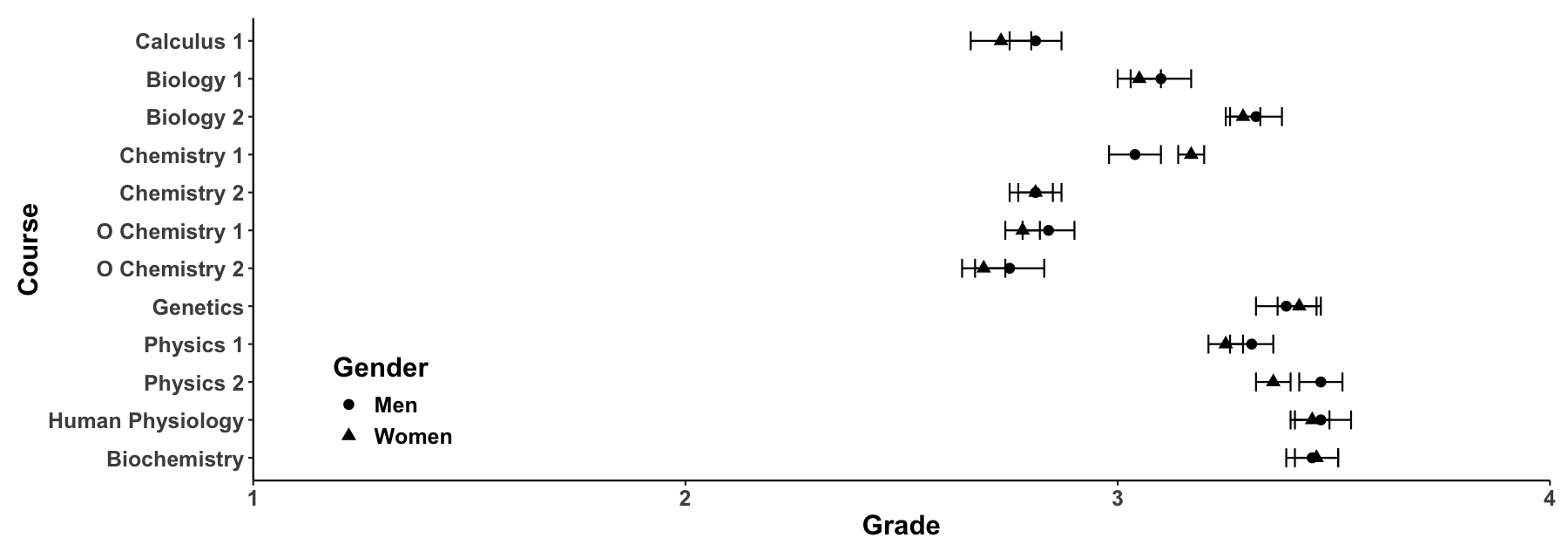}}}\vspace{30pt}
\subfloat[ \label{fig:grade_gender_inperson}]{%
{\includegraphics[width=1.0\textwidth]{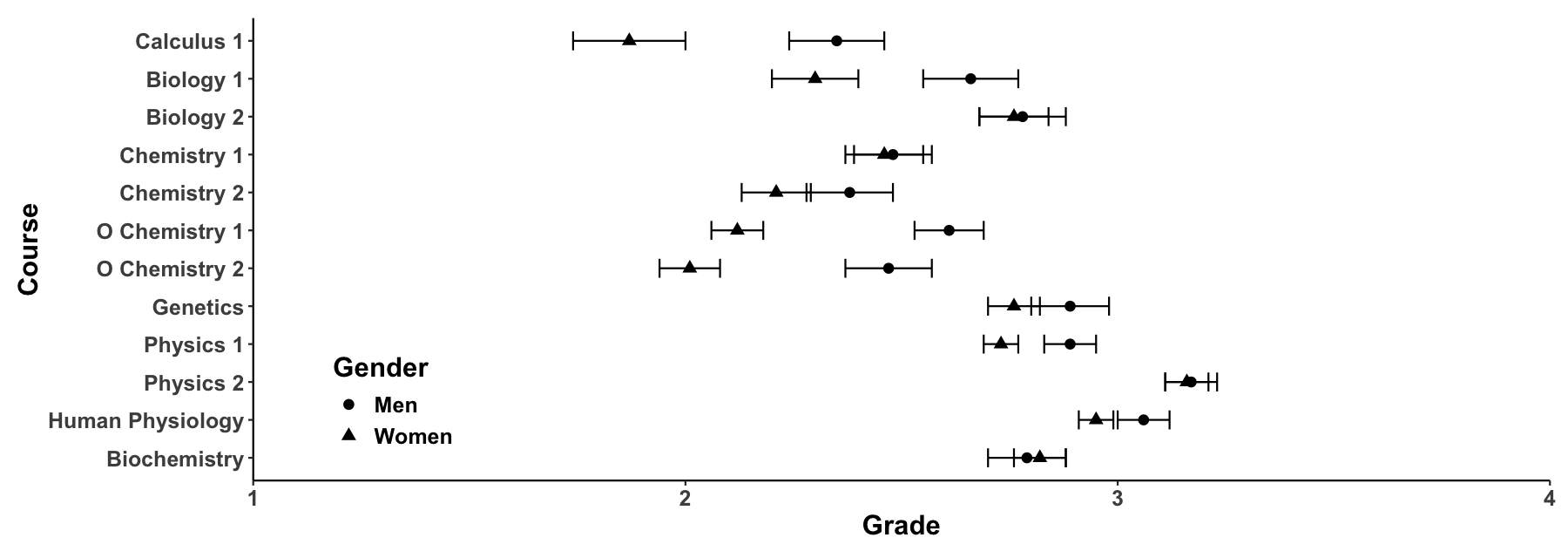}}}
\caption{Comparison of student grades for men and women for each course of interest for classes before (a), during (b), and after (c) remote instruction due to the COVID-19 pandemic. Ranges represent standard error of the mean.} \label{fig:grade_gender}
\end{figure}

There were generally more gender differences in AGAs than in grades.  For example, before remote instruction, there were statistically significant gendered grade differences in two courses (Calculus 1 and Genetics, see Figure \ref{fig:grade_gender}a), in which women tended to have higher grades than men. The statistical significance levels for the gendered differences in each course and period are shown in Tables \ref{biology_table}, \ref{chemistry_table}, \ref{math_table}, and \ref{physics_table}. During remote instruction, there was only one gendered grade difference (in Introductory Chemistry 1, favoring women). During post-remote instruction, all gendered grade differences (in Calculus 1, Biology 1, Organic Chemistry 1 and 2, and Physics 1, which can be seen in \mbox{Figure \ref{fig:grade_gender}c)} favored men. However, most courses did not have gendered grade differences. This trend is different from gendered AGA differences.  Before the pandemic, men had smaller grade penalties than women in six of the twelve of the courses studied, while women had smaller grade penalties in only one course, Genetics. During remote instruction, men again had smaller grade penalties than women in six courses and, after remote instruction, men had smaller grade penalties than women in seven courses. There were no courses in which women had smaller grade penalties than men during remote and post-remote  instruction. However, in most first-year courses (refer to Table \ref{semester}), women had larger grade penalties than men. Below, we investigate trends in specific subjects regarding student grades \mbox{and AGAs.}

\begin{figure}
\centering
\subfloat[\label{fig:aga_combo_precovid}]{%
{\includegraphics[width=1.0\textwidth]{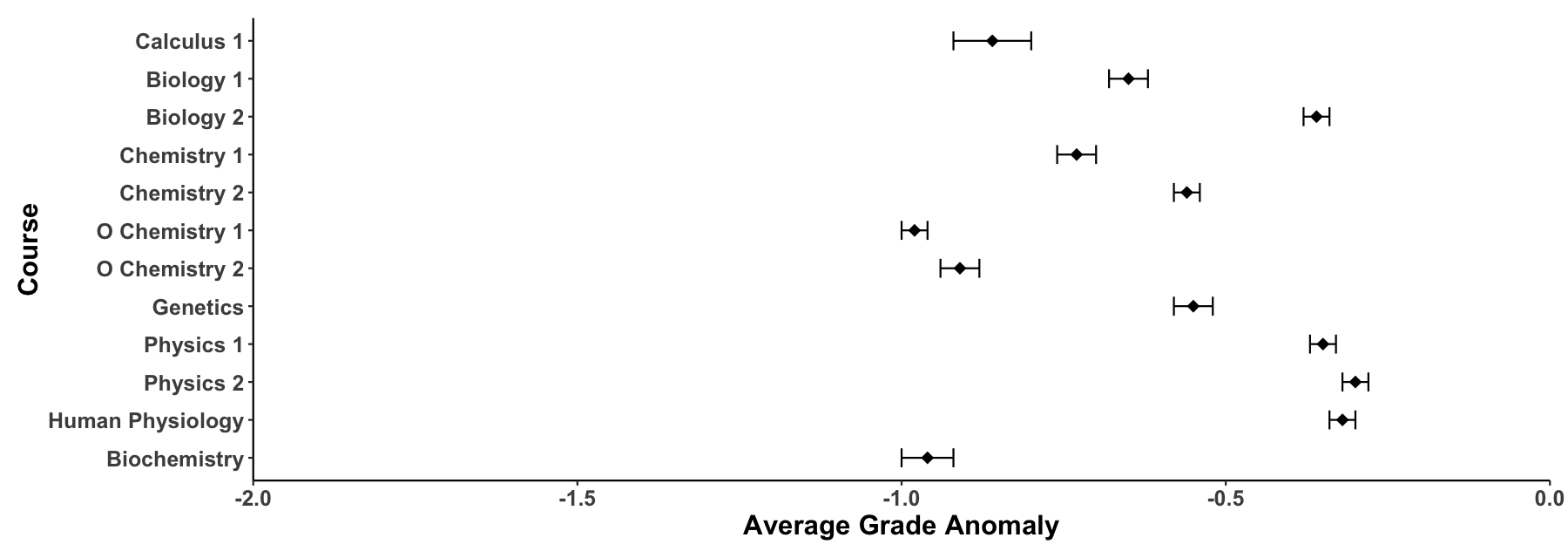}}}\vspace{30pt}
\subfloat[ \label{fig:aga_combo_remote}]{%
{\includegraphics[width=1.0\textwidth]{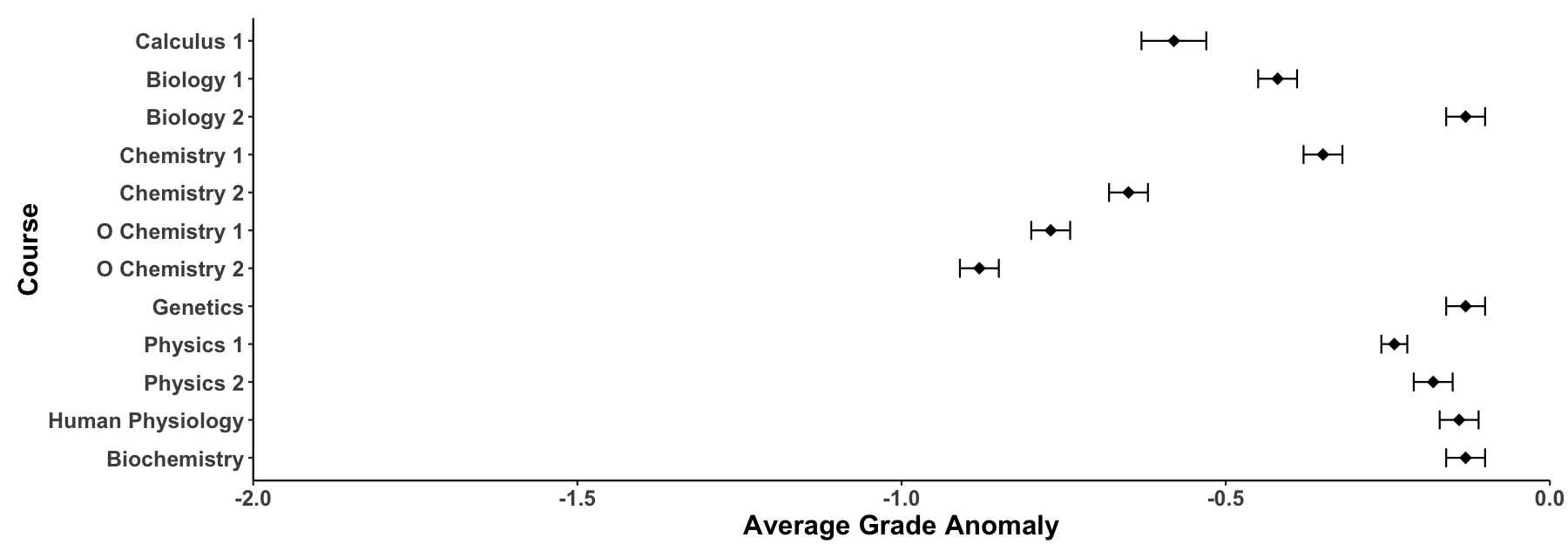}}}\vspace{30pt}
\subfloat[ \label{fig:aga_combo_inperson}]{%
{\includegraphics[width=1.0\textwidth]{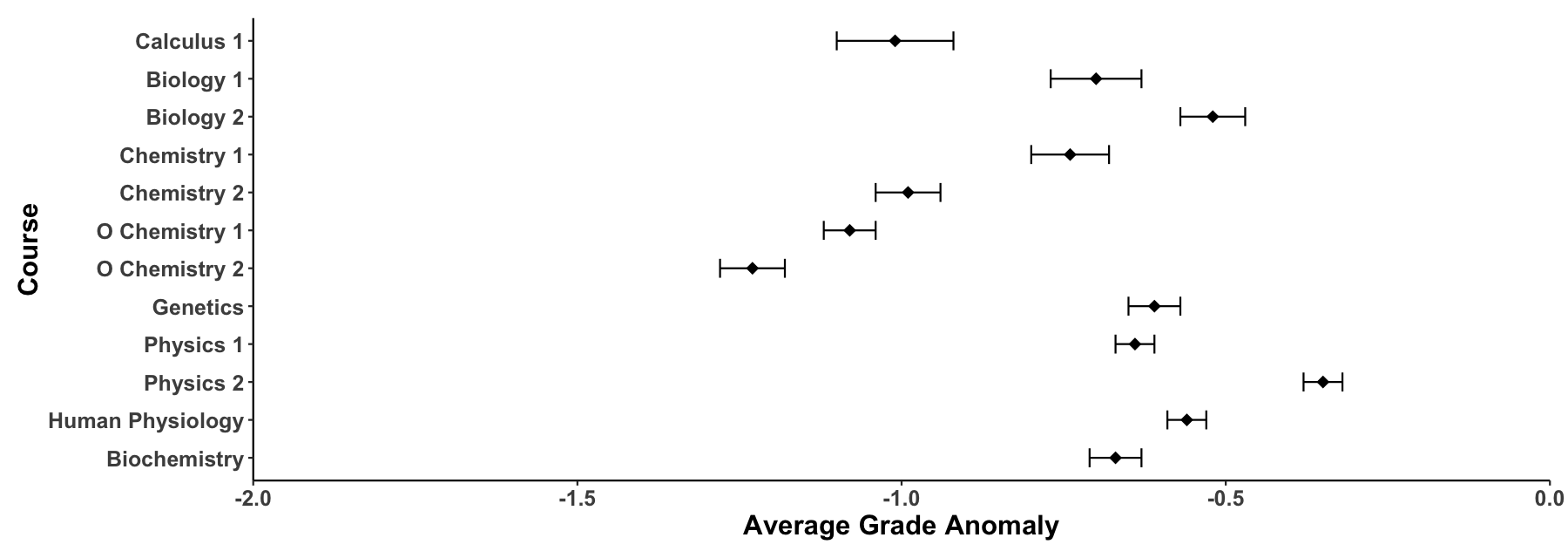}}}
\caption{Comparison of student average grade anomalies for each course of interest for classes before (a), during (b), and after (c) remote instruction due to the COVID-19 pandemic. Ranges represent standard error of the mean.} \label{fig:aga_combo}
\end{figure}

\begin{figure}
\centering
\subfloat[ \label{fig:aga_gender_precovid}]{%
{\includegraphics[width=1.0\textwidth]{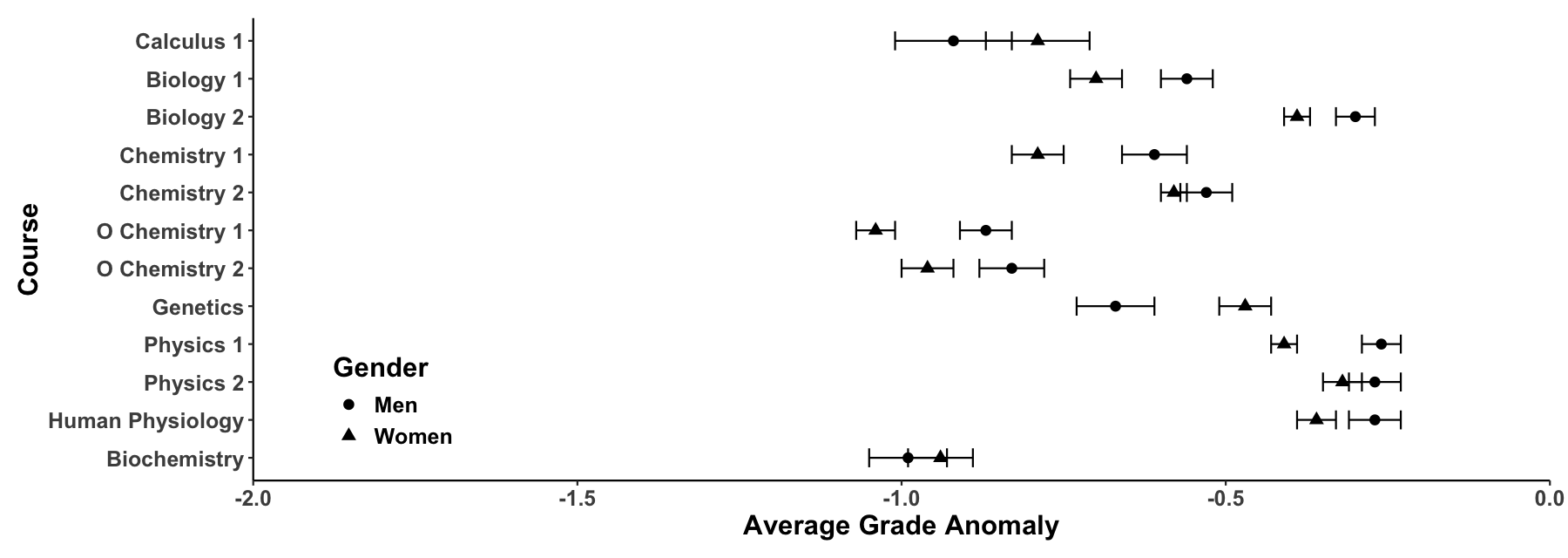}}}\vspace{30pt}
\subfloat[ \label{fig:aga_gender_remote}]{%
{\includegraphics[width=1.0\textwidth]{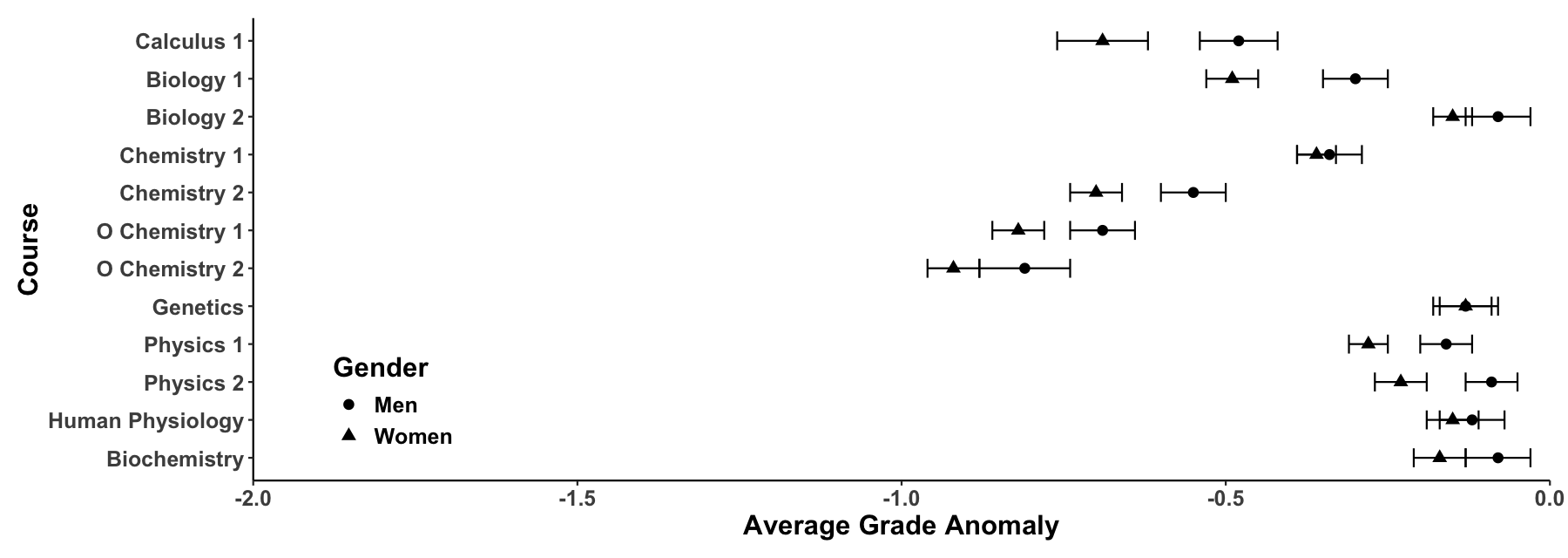}}}\vspace{30pt}
\subfloat[ \label{fig:agaa_gender_inperson}]{%
{\includegraphics[width=1.0\textwidth]{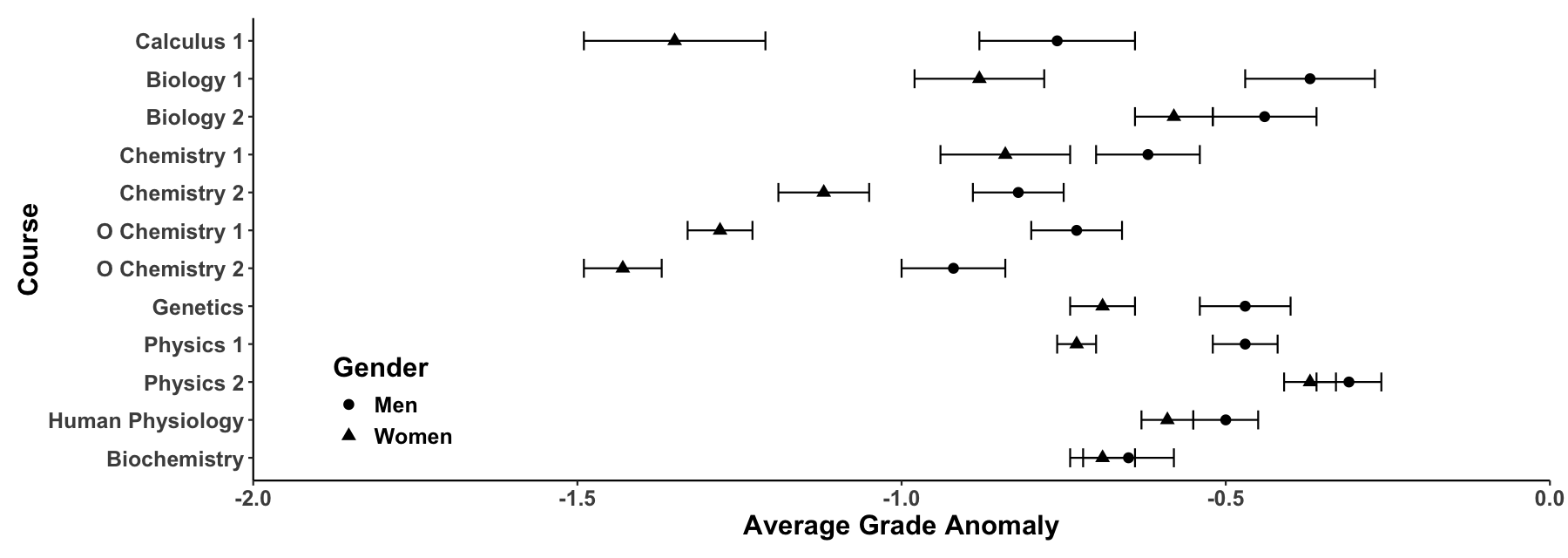}}}
\caption{Comparison of student average grade anomalies for men and women for each course of interest for classes before (a), during (b), and after (c) remote instruction due to the COVID-19 pandemic. Ranges represent standard error of the mean.} \label{fig:aga_gender}
\end{figure}

\subsection{Biological Science Courses}

Grades for Biological Science courses, including Biology 1 and 2, Genetics, Human Physiology, and Biochemistry, were generally higher compared to courses offered by the other departments, before, during, and after remote instruction. This means that, among the courses included in this study, students in bioscience and health-related majors tended to have their highest grades in courses offered by the Biological Sciences Department, which can be seen in Figure \ref{fig:grade_combo}. Biological Science grades during remote instruction were higher than during pre- or post-remote instruction. Post-remote course grades tended to be the same or one fraction of a  letter grade lower than pre-remote grades with the exception of Biochemistry (for example, in Biology 1 the average course grade was B$-$ before remote instruction, but C$+$ after, which can be seen in Table \ref{grade_combo_table}). 

However, some Biological Science classes had gender differences in grades, which can be seen in Table \ref{biology_table}. Genetics had the largest gender difference in course grades ($d=0.26$) among the courses studied pre-remote instruction (favoring women), and Biology 1 had a statistically significant gender difference in course grades ($d=-0.32$) post-remote instruction (favoring men). 

AGAs for Biological Science courses also tended to be average or small compared to other courses studied, as shown in Figure \ref{fig:aga_combo}. Although Bioscience students can generally expect lower grades in Biological Science courses compared to most courses they take, the grade penalties are relatively modest compared to other STEM courses included in this study.  Biological Science course AGAs tended to have the smallest magnitude during remote instruction, as seen in Figure \ref{fig:aga_combo}b and Table \ref{AGA_combo_table} in the appendix. AGAs tended to be either similar or slightly larger in magnitude after post-remote than pre-remote instruction. One exception to this trend was Biochemistry, which had a smaller grade penalty during post-remote than during pre-remote instruction. 

Most Biological Science courses had gender differences in AGAs during at least one period of the study (see Table \ref{biology_table}). Most of these gender differences showed that women had larger grade penalties on average than men with the exception of Genetics during pre-remote instruction.  Biology 1 and Biology 2 had small ($d$$\sim$0.2) gender differences pre-remote teaching, and Biology 1 had a small gender difference during remote instruction. Genetics did not have any gender differences during remote instruction, but had a small gender difference during post-remote instruction favoring men. Biology 1 grade anomaly gender differences became larger over time, and during post-remote instruction there was a medium grade difference ($d$$\sim$0.5) between men and women, favoring men.

\begin{table}[tb]
\tbl{Means and standard deviations (SD) of grades and grade anomalies by gender for courses offered by the Biological Science Department. \label{biology_table}}
{\begin{tabular}{ll|ccccc|ccccc|cc} \toprule
 &  & \multicolumn{5}{c}{Women} & \multicolumn{5}{c}{Men} & & \\ \cmidrule{3-7} \cmidrule{8-12}
 & &  & \multicolumn{2}{c}{AGA} & \multicolumn{2}{c}{Grade} & & \multicolumn{2}{c}{AGA} & \multicolumn{2}{c}{Grade} & \multicolumn{2}{c}{Cohen's \textit{d}} \\
    Course & Type & N & Mean & SD & Mean & SD & N & Mean & SD & Mean & SD & AGA & Grade \\ \midrule
    Biology 1 & Pre-Remote & 1077 & -0.70 & 1.17 & 2.86 & 1.02 & 571 & -0.55 & 0.92 & 2.88 & 1.00 & \textbf{-0.13$^{*}$} & -0.02  \\ 
    & Remote & 305 & -0.49 & 0.74 & 3.05 & 0.81 & 166 & -0.30 & 0.70 & 3.10 & 0.88 & \textbf{-0.26$^{**}$} & -0.05  \\ 
    & Post-Remote & 139 & -0.88 & 1.15 & 2.30 & 1.19 & 73 & -0.37 & 0.87 & 2.66 & 0.98 & \textbf{-0.47$^{**}$} &\textbf{-0.32$^{*}$}  \\ \midrule
    Biology 2 & Pre-Remote& 898 & -0.39 & 0.65 & 3.09 & 0.65 & 517 & -0.30 & 0.63 & 3.11 & 0.79 & \textbf{-0.13$^{*}$} & -0.03  \\ 
    & Remote & 352 & -0.15 & 0.64 & 3.29 & 0.78 & 172 & -0.08 & 0.62 & 3.32 & 0.76 & -0.12 & -0.03  \\
    & Post-Remote & 112 & -0.58 & 0.64 & 2.76 & 0.82 & 90 & -0.44 & 0.73 & 2.78 & 0.94 & -0.19 & -0.02  \\ \midrule
    Genetics & Pre-Remote & 449 & -0.47 & 0.76 & 3.02 & 0.94 & 307 & -0.67 & 1.00 & 2.75 & 1.18 &\textbf{0.23$^{**}$} & \textbf{0.26$^{***}$} \\ 
    & Remote& 229 & -0.13 & 0.61 & 3.42 & 0.74 & 135 & -0.13 & 0.61 & 3.39 & 0.77 & 0.00 & 0.04  \\
    & Post-Remote & 241 & -0.69 & 0.78 & 2.76 & 0.98 & 136 & -0.47 & 0.78 & 2.89 & 1.08 & \textbf{-0.28$^{**}$} & -0.13  \\ \midrule
    Human Physiology & Pre-Remote & 568 & -0.36 & 0.76 & 3.18 & 0.90 & 333 & -0.27 & 0.71 & 3.20 & 0.87 & -0.12 & -0.03  \\ 
    & Remote & 262 & -0.15 & 0.61 & 3.45 & 0.71 & 145 & -0.12 & 0.66 & 3.47 & 0.80 & -0.04 & -0.03  \\
    & Post-Remote & 516 & -0.60 & 0.80 & 2.95 & 0.98 & 283 & -0.50 & 0.82 & 3.06 & 1.00 & -0.11 & -0.10  \\ \midrule 
    Biochemistry & Pre-Remote &  311 & -0.94 & 0.86 & 2.59 & 1.04 & 212 & -0.99 & 0.91 & 2.42 & 1.13 & 0.06 & 0.15 \\  
    & Remote & 241 & -0.17 & 0.65 & 3.46 & 0.76 & 158 & -0.08 & 0.67 & 3.45 & 0.80 & -0.14 & 0.01 \\ 
    & Post-Remote & 371 & -0.69 & 1.00 & 2.82 & 1.22 & 193 & -0.65 & 1.02 & 2.79 & 1.29 & -0.03 & 0.03 \\ \bottomrule
    \end{tabular}}
    \raggedright
\bigskip 
\small\textit{Note}. Cohen's \textit{d} is positive if women had higher grades or smaller AGAs than men in a course. A bold Cohen's \textit{d} signifies that a \textit{t}-test showed significant differences between men and women.\\
$^{*} = p < 0.05$, $^{**} = p < 0.01$, and $^{***} = p < 0.001$.
\end{table}

\subsection{Chemistry Courses}

Chemistry courses tended to have some of the lowest grades among any courses studied, shown in Figure \ref{fig:grade_combo}. Notably, Organic Chemistry 2 had the lowest average grade of the courses studied during remote instruction. Figure \ref{fig:grade_combo} also shows that Organic Chemistry 1 had similar grades to Organic Chemistry 2, though they were slightly higher than those in Organic Chemistry 2 for remote and post-remote instruction. Introductory Chemistry 1 and 2 had grades that were either about or below the average of all courses. This means that, among the courses included in this study, students with bioscience and health-related majors tended to have their lowest grades in courses offered by the Chemistry Department.  Mostly, grades during remote instruction were higher than before or after, as seen in Figure \ref{fig:grade_combo}.  Course grades after remote instruction tended to be the same or one partial letter grade lower than they were before remote instruction.  Before remote instruction, Table \ref{chemistry_table} shows that no Chemistry courses had statistically significant gender differences in grades. However, during remote instruction, women had higher grades in Introductory Chemistry 1 and, during post-remote instruction, men tended to have higher grades in both Organic Chemistry 1 and 2. 

AGAs for Chemistry courses also tended to be large compared to other courses studied, as shown in Figure \ref{fig:aga_combo}.  Chemistry course grade anomalies tended to have the smallest magnitude during remote instruction. AGAs tended to be either similar or slightly larger during post-remote instruction than pre-remote instruction, which can be seen in \mbox{Table \ref{AGA_combo_table}} in  the appendix.
 Notably, during post-remote instruction, students taking Organic Chemistry 1 or 2 could expect to receive a grade over one full letter grade lower than their GPA excluding these classes. 

Men had smaller grade anomalies in all Chemistry classes (see Table \ref{chemistry_table}). All courses except for Introductory Chemistry 2 had a small but statistically significant gender difference in AGA during pre-remote instruction. Introductory Chemistry 2 and Organic Chemistry 1 had a small gender difference in AGA during remote instruction which grew larger after post-remote instruction.  Men tended to have smaller grade penalties than women with a medium effect size ($d$$\sim$0.5) in both Organic Chemistry 1 and 2 for post-remote instruction.

\begin{table}[tb]
\tbl{Means and standard deviations (SD) of grades and grade anomalies by gender for courses offered by the Chemistry Department.  \label{chemistry_table}}
{\begin{tabular}{ll|ccccc|ccccc|cc} \toprule
 &  & \multicolumn{5}{c}{Women} & \multicolumn{5}{c}{Men} & & \\ \cmidrule{3-7} \cmidrule{8-12}
 & &  & \multicolumn{2}{c}{AGA} & \multicolumn{2}{c}{Grade} & & \multicolumn{2}{c}{AGA} & \multicolumn{2}{c}{Grade} & \multicolumn{2}{c}{Cohen's \textit{d}} \\
    Course & Type & N & Mean & SD & Mean & SD & N & Mean & SD & Mean & SD & AGA & Grade \\ \midrule
    Introductory Chemistry 1 & Pre-Remote & 1166 & -0.79 & 1.38 & 2.86 & 0.87 & 677 & -0.61 & 1.41 & 2.82 & 0.97 &\textbf{-0.13$^{**}$}  & 0.04  \\ 
    & Remote & 358 & -0.36 & 0.64 & 3.17 & 0.66 & 191 & -0.34 & 0.69 & 3.04 & 0.78 & -0.03 & \textbf{0.19$^{*}$}  \\ 
    & Post-Remote & 133 & -0.83 & 1.12 & 2.46 & 1.06 & 122 & -0.62 & 0.88 & 2.48 & 0.97 & -0.21 & -0.02  \\ \midrule
    Introductory Chemistry 2 & Pre-Remote & 973 & -0.58 & 0.76 & 2.94 & 0.86 & 602 & -0.53 & 0.86 & 2.85 & 0.91 & -0.06 & 0.10  \\
    & Remote & 352 & -0.70 & 0.69 & 2.81 & 0.82 & 203 & -0.44 & 0.65 & 2.81 & 0.87 & \textbf{-0.22$^{*}$}  & 0.00  \\ 
    & Post-Remote & 171 & -1.12 & 0.92 & 2.21 & 1.08 & 134 & -0.82 & 0.83 & 2.38 & 1.10 & \textbf{-0.35$^{**}$} & -0.16  \\ \midrule
    Organic Chemistry 1 & Pre-Remote & 1058 & -1.04 & 0.99 & 2.47 & 1.12 & 589 & -0.87 & 0.94 & 2.57 & 1.07 & \textbf{-0.18$^{***}$} & -0.10  \\ 
    & Remote & 497 & -0.82 & 0.79 & 2.78 & 0.91 & 244 & -0.69 & 0.71 & 2.84 & 0.88 & \textbf{-0.17$^{*}$} & -0.07  \\
    & Post-Remote & 331 & -1.28 & 0.95 & 2.12 & 1.12 & 191 & -0.73 & 0.97 & 2.61 & 1.14 &\textbf{-0.57$^{***}$} &\textbf{-0.44$^{***}$} \\  \midrule
    Organic Chemistry 2 & Pre-Remote & 687 & -0.96 & 0.92 & 2.55 & 1.07 & 420 & -0.83 & 0.98 & 2.64 & 1.17 & \textbf{-0.13$^{*}$} & -0.08  \\ 
    & Remote  & 411 & -0.91 & 0.81 & 2.69 & 0.98 & 197 & -0.81 & 0.93 & 2.75 & 1.09 & -0.13 & -0.05  \\
    & Post-Remote & 269 & -1.43 & 1.00 & 2.01 & 1.19 & 175 & -0.92 & 1.07 & 2.47 & 1.27 & \textbf{-0.50$^{***}$} & \textbf{-0.38$^{***}$}  \\ \bottomrule
    \end{tabular}}
    \raggedright
\bigskip 
\small\textit{Note}. Cohen's \textit{d} is positive if women had higher grades or smaller AGAs than men in a course. A bold Cohen's \textit{d} signifies that a \textit{t}-test showed significant differences between men and women. \\
$^{*} = p < 0.05$, $^{**} = p < 0.01$, and $^{***} = p < 0.001$.
\end{table}

\subsection{Math Courses}
Table \ref{grade_combo_table} in the appendix shows that, during pre-remote instruction, Calculus 1 had an average grade of 2.45. During remote classes, Calculus 1 had an average grade of 2.77. During post-remote classes, the average grade was 2.15, which is the lowest average grade of any course studied across all periods.  Calculus 1 had statistically significant gender differences in grades during pre-remote instruction, in which women tended to have higher grades than men; however, this trend reversed after post-remote instruction. Calculus 1 AGAs were $-$0.86 before, $-$0.58 during, and $-$1.01 {after remote} instruction. This means that, for post-remote instruction, students were likely to have a Calculus grade more than one full letter grade lower than their GPA in other courses. There were no gender differences in AGAs before the pandemic, but women had larger AGAs than men during remote and post-remote instruction. 

\begin{table}[tb]
\tbl{Means and standard deviations (SD) of grades and grade anomalies by gender for the course offered by the Mathematics Department.  \label{math_table}}
{\begin{tabular}{ll|ccccc|ccccc|cc} \toprule
 &  & \multicolumn{5}{c}{Women} & \multicolumn{5}{c}{Men} & & \\ \cmidrule{3-7} \cmidrule{8-12}
 & &  & \multicolumn{2}{c}{AGA} & \multicolumn{2}{c}{Grade} & & \multicolumn{2}{c}{AGA} & \multicolumn{2}{c}{Grade} & \multicolumn{2}{c}{Cohen's \textit{d}} \\
    Course & Type & N & Mean & SD & Mean & SD & N & Mean & SD & Mean & SD & AGA & Grade \\ \midrule
    Calculus 1 & Pre-Remote & 385 & -0.79 & 1.48 & 2.58 & 1.11 & 436 & -0.92 & 1.90 & 2.34 & 1.30 & 0.07 & \textbf{0.20$^{**}$}  \\  
    & Remote & 184 & -0.69 & 0.94 & 2.73 & 0.93 & 179 & -0.48 & 0.84 & 2.81 & 0.87 & \textbf{-0.24$^{*}$} & -0.08 \\ 
    & Post-Remote & 123 & -1.35 & 1.53 & 1.87 & 1.41 & 165 & -0.76 & 1.51 & 2.35 & 1.36 & \textbf{-0.39$^{**}$} & \textbf{-0.35$^{**}$}  \\
    \bottomrule
    \end{tabular}}
    \raggedright
\bigskip 
\small\textit{Note}. Cohen's \textit{d} is positive if women had higher grades or smaller AGAs than men in a course. A bold Cohen's \textit{d} signifies that a \textit{t}-test showed significant differences between men and women. \\
$^{*} = p < 0.05$, $^{**} = p < 0.01$, and $^{***} = p < 0.001$.
\end{table}

\subsection{Physics Courses}

Before, during, and after remote instruction, Physics course grades were generally higher than most of the other subjects, which can be seen in Figure \ref{fig:grade_combo}. For both Physics 1 and 2, Table \ref{grade_combo_table} in the appendix shows that grades were highest during remote instruction.  For Physics 1, the average grade during pre-remote instruction was a B, but dropped to a B$-$ during post-remote instruction. For Physics 2, the average grade during both pre- and post-remote instruction was close to a B$+$. Table \ref{physics_table} shows that neither Physics 1 nor 2 had gendered grade differences during pre-remote or remote instruction. However, during post-remote instruction for Physics 1, men tended to have higher grades than women, with a small effect size ($d$$\sim$0.2). 

AGAs in Physics courses also tended to have a smaller magnitude than those of other subjects (see Figure \ref{fig:aga_combo}) and for both Physics courses AGAs were smallest during remote instruction. Physics 1 courses had small ($d$$\sim$0.2) AGA gender differences favoring men before, during, and after remote instruction. Physics 2 had no statistically significant gender difference in AGA before or after remote instruction, and men tended to have smaller grade penalties than women during remote instruction, with a small effect size ($d$$\sim$0.2). 

\begin{table}[tb]
\tbl{Means and standard deviations (SD) of grades and grade anomalies by gender for courses offered by the Physics Department.  \label{physics_table}}
{\begin{tabular}{ll|ccccc|ccccc|cc} \toprule
 &  & \multicolumn{5}{c}{Women} & \multicolumn{5}{c}{Men} & & \\ \cmidrule{3-7} \cmidrule{8-12}
 & &  & \multicolumn{2}{c}{AGA} & \multicolumn{2}{c}{Grade} & & \multicolumn{2}{c}{AGA} & \multicolumn{2}{c}{Grade} & \multicolumn{2}{c}{Cohen's \textit{d}} \\
    Course & Type & N & Mean & SD & Mean & SD & N & Mean & SD & Mean & SD & AGA & Grade \\ \midrule
    Physics 1 & Pre-Remote & 732 & -0.41 & 0.60 & 3.06 & 0.76 & 435 & -0.26 & 0.63 & 3.13 & 0.84 & \textbf{-0.25$^{***}$} & -0.09  \\
    & Remote & 390 & -0.28 & 0.60 & 3.24 & 0.75 & 247 & -0.16 & 0.62 & 3.31 & 0.80 & \textbf{-0.20$^{*}$} & -0.09  \\ 
    & Post-Remote & 601 & -0.73 & 0.78 & 2.73 & 0.94 & 292 & -0.47 & 0.77 & 2.89 & 0.98 & \textbf{-0.33$^{***}$}& \textbf{-0.16$^{*}$} \\  \midrule
    Physics 2 & Pre-Remote & 457 & -0.32 & 0.61 & 3.19 & 0.77 & 289 & -0.27 & 0.67 & 3.15 & 0.87 & -0.07 & 0.04  \\  
    & Remote & 330 & -0.23 & 0.67 & 3.36 & 0.73 & 197 & -0.09 & 0.53 & 3.47 & 0.71 & \textbf{-0.24$^{**}$} & -0.16  \\
    & Post-Remote & 374 & -0.37 & 0.73 & 3.16 & 0.91 & 249 & -0.31 & 0.74 & 3.17 & 0.99 & -0.09 & -0.01  \\ 
    \bottomrule
    \end{tabular}}
    \raggedright
\bigskip 
\small\textit{Note}. Cohen's \textit{d} is positive if women had higher grades or smaller AGAs than men in a course. A bold Cohen's \textit{d} signifies that a \textit{t}-test showed significant differences between men and women.\\
$^{*} = p < 0.05$, $^{**} = p < 0.01$, and $^{***} = p < 0.001$.
\end{table}

\section{Discussion}

\subsection{ Do Grades or Grade Anomalies Differ before, during, and after Remote Instruction due to the COVID-19 Pandemic?}

Grades are important to students for a variety of reasons such as continuing their major, scholarship requirements, and graduate school or professional school admissions.  Broadly, grades were higher during remote instruction and were lower again during post-remote instruction.  AGAs, unlike grades, do not have a direct effect on students' outcomes such as scholarships and graduate admissions. A student with an A average who receives a B in a class has the same grade anomaly as a student with a B average who receives a C in the class. Here, we use the idea of academic self-concept from Situated Expectancy Value Theory to frame how students may think about grade anomalies \cite{eccles2020}. Students tend to have a somewhat fixed idea of what sort of student they are (for example, they may endorse the idea that ``If I get As, I must be an A kind of person'') \cite{seymour2019}. Grade anomalies may challenge a student's idea about what kind of student they are. In particular,  students may compare their grades across courses to determine which disciplines they excel at or struggle with \cite{eccles2020}. 

During remote instruction, there tended to be increases in average grades compared to pre-remote instruction for most of the classes. These increases in grades may be due to a range of factors. For example, grading schemes and assessment types may have been changed, or instructors may have been more flexible with grading than during pre-remote classes \cite{chan2022}. We note that student performance on content-based surveys is similar in online and in-person administration \cite{vandusen2021}, and answer copying on homework does not significantly differ between remote and in-person instruction \cite{chen2022}. This leaves open the possibility that grade differences between in-person and online courses are not inherent, but may be the result of instructor choices in class policies. 

During post-remote instruction, average grades for all courses were lower compared to remote instruction and AGAs were larger in magnitude. That is, students' grades were more consistent during remote instruction, so that for most classes the grades deviated less from a students' average GPA during remote instruction. We hypothesize that these smaller grade anomalies may result in students being less concerned that they can succeed in their discipline, and may rely more on other factors (such as interest) to make decisions regarding major and career choice.  

Comparing grades for pre- and post-remote instruction, we can see that the average was lower for all courses during post-remote instruction, except for Biochemistry. Similarly, there was no course in which AGAs decreased in magnitude from {pre- to post-remote instruction} 
except for Biochemistry, and almost all courses had larger AGAs during post-remote than \mbox{pre-remote courses.} 

The courses with the lowest grades and largest magnitude of AGAs over all time periods were Calculus 1 and Organic Chemistry 1 and 2, which are often labeled as ``weed out'' courses. Grade penalties are more common and larger in STEM disciplines than in social sciences or humanities  \cite{seymour2019, rask2010, koester2016, matz2017}, but our findings show that there are significant variations in AGAs even among STEM courses. In general, Biological Science and Physics courses tended to have had the smallest AGAs, while Math and Chemistry courses tended to have the largest. Thus, this is not a simple issue of STEM courses having larger AGAs than non-STEM courses.  Instructors and departments with comparatively lower grades and larger AGAs than others may benefit from pedagogies implemented by other STEM departments and instructors at their institution. 

A variety of factors are likely contributing to the changes in grades and AGAs over time. Although it is possible that some students may be cheating, cheating on exams seemed to have only small increases in the USA during the pandemic, though the effect may be larger in other regions \cite{ives2023}. Also, research suggests that certain factors might have led to improved grades during remote instruction. For example, because there were more low-stakes assessments during remote instruction, students may have been more likely to engage in spaced practice instead of ``cramming'' for assessments during remote learning \cite{gonzalez2020}.  One study showed that students had higher grades during COVID-19 remote instruction even on identical assessments that were also given online pre-pandemic \cite{gonzalez2020}. Another study that focused on quantum mechanics (an upper-level physics course) found that implementing low-stakes formative assessments instead of exams did not lead to lower scores on the course post-test (which only contributed a small amount to the students' final grade) \cite{palmgren2022}. While these studies do not pinpoint specific reasons for differences in grades between remote and in-person classes, they do suggest that increases in grades do not necessarily correlate with lowered academic standards or cheating.

One of the concerning trends from this study was that there are generally lower grades and larger AGAs for students in their first two years of university than for later years (see Table \ref{semester} in the appendix for information about when students tend to take each course). Low grades early on during the transition to university are particularly concerning. Low grades, even if they are high enough to continue in a major, are a common reasons that students cite for leaving a STEM major \cite{seymour1997,seymour2019}.  Additionally, because academic self-concept is most in flux during transitional periods, low grades early in a student's college career may negatively affect students' self-concept more than low grades received in later \mbox{years \cite{gniewosz2012}.}

\subsection{Are There Gender Differences in Grades or Grade Anomalies, and Do They Differ before, during, and after Remote Instruction due to the COVID-19 Pandemic?}

Before remote instruction, there were no statistically significant grade differences between men and women, except for Genetics and Calculus 1 where women had higher grades than men. However, for remote and post-remote instruction, all classes with statistically significant grade differences (except for Introductory Chemistry 1 during remote instruction) favored men. In each case, there were only a few courses with any statistically significant grade differences.  One particularly concerning class was Calculus 1. During pre-remote and remote instruction, women had an average grade which was slightly over C$+$, but an average grade below C during post-remote instruction. This means that, on average, women did not have a grade in Calculus 1 needed to continue in the major. These women who failed this course needed to choose between taking the class again, which involves a commitment of both time and tuition, or change to a major that does not require Calculus. Although other courses had similar or larger gender differences in grades, this is the only course and group for which the average outcome was not passing the course (i.e., have that course count towards their major). 

There were many courses that had statistically significant gendered AGA differences. In all of these courses (except pre-remote Genetics), men had smaller AGAs than women.  The largest of these AGA differences were post-remote Biology 1, Organic Chemistry 1, and Organic Chemistry 2. For women in bioscience majors, a large grade anomaly in their first Biological Science course at university may be particularly concerning, and potentially lead them to believe that they do not ``have what it takes'' to succeed in their major.  Women often report worrying more than men that they do not understand the material even if they receive As, Bs, or Cs (which are grades that allow students to continue in most programs)~\cite{goodman2002, seymour2019}. This trend has been found to be particularly strong among high-achieving women~\cite{seymour2019}. 

We hypothesize that women may experience a lower academic self-concept compared to men at similar performance levels. Previous research suggests that men are more likely to differentiate between their grades and their sense of academic self-concept~\mbox{\cite{seymour1997, seymour2019, rask2008}.} Academic self-concept is formed through grades and feedback from outsiders. Women are generally less likely to receive recognition from instructors \cite{eaton2020,mossracusin2012,wang2018}; therefore, women might depend more on grade information than men to shape their academic self-concept~\mbox{\cite{seymour1997, seymour2019, rask2008}.} Women also tend to earn higher grades than men who have the same standardized test scores \cite{seymour2019, voyer2014}, so they may be more accustomed to higher grades. As a result, they may have more concern about grades that are lower than what they are accustomed to, or they may compare their relatively low STEM grades and view themselves as less able to succeed in biological sciences or health-related fields than a subject that gives them the recognition for their work that they are accustomed to \cite{seymour1997,seymour2019}.

As seen in our results, AGAs and raw grade data do not necessarily align at all times. There are more gender differences in AGAs than in grades in the findings presented here. This highlights the importance of using AGA as a measure. For example, an instructor might not see any gender disparities in grades; however, without knowing the gender differences in AGA, they may not recognize how those grades may be perceived by women and men in their classes. By looking at both grades and AGAs, instructors can better identify and understand potential inequities in their classrooms.

\section{Conclusion and Future Research}

In this work, we found that grade penalties exist for all the courses taken by bioscience major students in the context of our study. Grades were higher and grade penalties were smaller during remote instruction compared to pre-remote instruction. During post-remote instruction, grades were lower and grade penalties were larger than during remote instruction. Furthermore, there were more gender differences in both grades and grade penalties (favoring men) during post-remote teaching than for pre-remote or remote teaching. 

We observed that the pattern of women experiencing larger grade penalties than men persists even among students in bioscience majors, where women are typically in the majority. This suggests that gender disparities in academic performance may persist across different fields and contexts, indicating systemic issues affecting women's academic self-concept regardless of their numerical representation. These results are also important because they provide evidence that courses in STEM departments tend to have grade penalties, and that these penalties tend to decrease during remote instruction.  Additionally, AGA may also act as a useful measure of academic self-concept that is easy for institutions to access. This study was conducted based on university-provided data and is quantitative in nature. We would like to note that, while we have identified trends and disparities in grades and grade anomalies among these students, future research would greatly benefit from incorporating qualitative methods, such as interviews with students, as they provide a deeper understanding of the experiences and perspectives behind the observed trends.

Although we have evidence of grade penalties in the studied courses as well as gendered grade anomaly differences, we did not have access to syllabi or other information about individual courses offered over the period of data collection. Therefore, we are unable to point out specific practices that may lead to grade penalties, grade bonuses, or gender inequities at this institution.

Finally, this research was conducted at a large, predominantly white public university. While our findings may be relevant to similar institutions, we do not have data on grade anomalies at smaller liberal arts colleges, minority-serving institutions, or community colleges in the US. Additionally, it would be valuable to replicate this research in other countries, as the impact of the COVID-19 pandemic varied globally.

\section*{Author Contributions}

A.M. and F.S. contributed to analysis and interpretation of data, as well as writing and revision of the manuscript. C.S. contributed to the conception and design of research, funding acquisition, and revision of the manuscript.

\section*{Funding}
This research is supported by National Science Foundation grant DUE-1524575.

\section*{Institutional Review}
This research was carried out in accordance with the principles outlined in this university's Institutional Review Board ethical policy.

\section*{Informed Consent} 

Consent requirement was waived due to the research being approved as exempt from informed consent by the university’s Institutional Review Board.

\section*{Data Availability} 

The data presented in this study are available on request from the corresponding author due to the data privacy requirements of US FERPA regulations.

\section*{Conflicts of Interest} 

No potential conflict of interest was reported by the authors.

\bibliographystyle{achemso} 
\bibliography{refs}

\newpage

\appendix
\renewcommand{\thetable}{\arabic{table}-A} 
\label{appendixA}
\section*{Appendix: Courses, Majors, and Composite grades for Women and Men}

\begin{table}[h]
\tbl{Course Requirements by Major\label{course_requirements}} 
{\begin{tabular}{l|cccccccccccc} \toprule
Major & Calc 1 & Bio 1 & Bio 2 &  Chem 1 & Chem 2 & Gen & OC 1 & OC 2 & HP & BC & Phys 1 & Phys 2  \\ \midrule
Biological Sciences & R & R & R & R & R & R & R & R & O & R & R & R\\
Computational Biology & R & R & R & R & R & R & R & &  & R & O &  \\
Ecology \& Evolution & R & R & R & R & R & R & R & R & & R & R & R \\
Microbiology & R & R & R & R & R & R & R & R & O & R & R & R \\
Molecular Biology & R & R & R & R & R & R & R & R & & & R & R  \\
Neuroscience & R & R & R & R & R & & R & R & R & R & R & R \\
Pharmacy & R & R & R & R & R & & R & R & & R & O & O \\
Rehabilitation Science & & R & & R & & & & & & & R & \\ \bottomrule
\end{tabular}}
\raggedright
\bigskip 
\small\textit{Note}. The abbreviations stand for: Calculus 1, Biology 1, Biology 2, Introductory Chemistry 1, Introductory Chemistry 2, Genetics, Organic Chemistry 1, Organic Chemistry 2, Human Physiology, Biochemistry, Physics 1, and Physics 2. R designates a required course, O designates a course that can be taken for elective credit in the major, and no letter designates a course that does not fulfill any credits for the major. 
\end{table}

\begin{table}[h]
\centering
\caption{Courses Studied by Year \label{semester}} 
{\begin{tabular}{ll|ccccc} \toprule
 Course & Course Type  & $1^{st}$ & $2^{nd}$ & $3^{rd}$ & $4^{th}$  & $\ge$ $5^{th}$  \\ \midrule
Calculus 1 & Pre-Remote & \textbf{68} & 21 & 6 & 4 & 1 \\
& Remote & \textbf{63} & 30 & 6 & 0 & 1 \\ 
& Post-Remote & \textbf{38} & 36 & 18 & 6 & 2 \\ \cmidrule{3-7}
Biology 1 & Pre-Remote & \textbf{90}  & 9  & 1  & 0 & 0 \\ 
& Remote & \textbf{82} & 16 & 1 & 1 & 0 \\ 
& Post-Remote & \textbf{69} & 20 & 7 & 2 & 2 \\ \cmidrule{3-7}
Biology 2 & Pre-Remote & \textbf{71}  & 24  & 4  & 1 & 0 \\ 
& Remote & \textbf{58} & 35 & 5 & 1 & 1 \\ 
& Post-Remote & 40 & \textbf{46} & 10 & 4 & 0 \\ \cmidrule{3-7}
Chemistry 1 & Pre-Remote & \textbf{93}  & 6  & 1  & 0  & 0  \\
& Remote & \textbf{86} & 12 & 2 & 0 & 0 \\ 
& Post-Remote & \textbf{69} & 21 & 8 & 1 & 1\\ \cmidrule{3-7}
Chemistry 2 & Pre-Remote & \textbf{77}  & 19  & 3  & 1  & 0    \\
& Remote & \textbf{65} & 28 & 6 & 0 & 1\\ 
& Post-Remote & \textbf{42} & 41 & 13 & 3 & 1 \\ \midrule
Organic Chemistry 1 & Pre-Remote & 5  & \textbf{81}  & 11  & 2  & 1   \\ 
& Remote & 6 & \textbf{85} & 7 & 2 & 0 \\ 
& Post-Remote & 8 & \textbf{68} & 20 & 3 & 1 \\ \cmidrule{3-7}
Organic Chemistry 2 & Pre-Remote & 2  & \textbf{63}  & 29  & 5  & 1  \\
& Remote & 3 & \textbf{71} & 19 & 5 & 2 \\ 
& Post-Remote & 4 & \textbf{52} & 34 & 8 & 2 \\ \cmidrule{3-7}
Genetics & Pre-Remote & 4  & \textbf{54}  & 33  & 8 & 1   \\ 
& Remote & 3 & \textbf{55} & 37 & 4 & 1 \\ 
& Post-Remote & 4 & 28 & \textbf{49} & 15 & 4 \\ \midrule
Physics 1 & Pre-Remote & 13  & 29  & \textbf{51}  & 6  & 1  \\ 
& Remote & 13 & 34 & \textbf{49} & 3 & 1 \\ 
& Post-Remote & 14 & 33 & \textbf{46} & 5 & 2\\  \cmidrule{3-7}
Physics 2 & Pre-Remote & 4  & 23  & \textbf{59}  & 12 & 2   \\
& Remote & 4 & 20 & \textbf{69} & 5 & 2 \\ 
& Post-Remote & 2 & 20 & \textbf{63} & 12 & 3\\ \cmidrule{3-7}
Human Physiology & Pre-Remote & 2  & 20  & \textbf{61}  & 15  & 2  \\
& Remote & 1 & 17 & \textbf{71} & 10 & 1 \\ 
& Post-Remote & 1 & 9 & \textbf{53} & 33 & 4 \\ \cmidrule{3-7}
Biochemistry & Pre-Remote & 1  & 8  & \textbf{62}  & 24  & 5  \\ 
& Remote & 1 & 10 & \textbf{80} & 6 & 3 \\ 
& Post-Remote & 0 & 8 & \textbf{63} & 22 & 7  \\
\bottomrule
\end{tabular}}\\
\raggedright
\bigskip 
\small\textit{Note}. List of courses studied, the period of interest, and the percentage of students in our sample who take each course in a given year. For example, 68\% of students take calculus during their first year of university, and 21\% of students take calculus during their second year. The year in which students most often take the course has its percentage of students in bold.
\end{table}

\begin{table}[h]
\tbl{Course grades before, during, and after remote instruction due to the COVID-19 pandemic.  \label{grade_combo_table}}
{\begin{tabular} {l|ccc|ccc|ccc} \toprule
 & \multicolumn{3}{c}{Pre-Remote} & \multicolumn{3}{c}{Remote} & \multicolumn{3}{c}{Post-Remote} \\
 Course & N & Mean & SD & N & Mean & SD &  N & Mean & SD \\ \midrule
        Calculus 1 & 821 & 2.45 & 1.22 & 363 & 2.77 & 0.90 & 288 & 2.15 & 1.40  \\ 
        Biology 1 & 1648 & 2.87 & 1.01 & 471 & 3.07 & 0.83 & 212 & 2.43 & 1.13  \\ 
        Biology 2 & 1415 & 3.10 & 0.78 & 524 & 3.30 & 0.77 & 202 & 2.77 & 0.87  \\ 
        Introductory Chemistry 1 & 1843 & 2.85 & 0.91 & 549 & 3.13 & 0.71 & 255 & 2.47 & 1.10  \\ 
        Introductory Chemistry 2 & 1575 & 2.90 & 0.88 & 555 & 2.81 & 0.84 & 305 & 2.28 & 1.09  \\ 
        Organic Chemistry 1 & 1647 & 2.50 & 1.10 & 741 & 2.80 & 0.90 & 522 & 2.30 & 1.15  \\ 
        Organic Chemistry 2 & 1107 & 2.58 & 1.11 & 608 & 2.71 & 1.02 & 444 & 2.19 & 1.24  \\ 
        Genetics & 756 & 2.91 & 1.05 & 364 & 3.41 & 0.75 & 377 & 2.80 & 1.02  \\ 
        Physics 1 & 1167 & 3.09 & 0.79 & 637 & 3.27 & 0.77 & 893 & 2.78 & 0.95  \\ 
        Physics 2 & 746 & 3.18 & 0.81 & 527 & 3.40 & 0.72 & 623 & 3.16 & 0.94  \\ 
        Human Physiology & 901 & 3.19 & 0.89 & 407 & 3.45 & 0.74 & 799 & 2.99 & 0.98  \\ 
        Biochemistry & 523 & 2.52 & 1.08 & 399 & 3.45 & 0.78 & 564 & 2.81 & 1.25 \\ \bottomrule
    \end{tabular}}
\raggedright
\bigskip 
\small\textit{Note}. Mean and standard deviation (SD) of average grades, as well as number of students (N) for each course of interest.
\end{table}

\begin{table}[h]
\tbl{Course grade anomalies before, during, and after remote instruction due to the COVID-19 pandemic.  \label{AGA_combo_table}}
{\begin{tabular} {l|ccc|ccc|ccc} \toprule
 & \multicolumn{3}{c}{Pre-Remote} & \multicolumn{3}{c}{Remote} & \multicolumn{3}{c}{Post-Remote} \\
 Course & N & Mean & SD & N & Mean & SD &  N & Mean & SD \\ \midrule
        Calculus 1 & 821 & -0.86 & 1.72 & 363 & -0.58 & 0.90 & 288 & -1.01 & 1.54  \\ 
        Biology 1 & 1648 & -0.65 & 1.10 & 471 & -0.42 & 0.73 & 212 & -0.70 & 1.09  \\ 
        Biology 2 & 1415 & -0.36 & 0.65 & 524 & -0.13 & 0.63 & 202 & -0.52 & 0.68  \\ 
        Introductory Chemistry 1 & 1843 & -0.73 & 1.40 & 549 & -0.35 & 0.66 & 255 & -0.74 & 1.02  \\ 
        Introductory Chemistry 2 & 1575 & -0.56 & 0.80 & 555 & -0.65 & 0.68 & 305 & -0.99 & 0.89  \\ 
        Organic Chemistry 1 & 1647 & -0.98 & 0.98 & 741 & -0.77 & 0.76 & 522 & -1.08 & 0.99  \\ 
        Organic Chemistry 2 & 1107 & -0.91 & 0.94 & 608 & -0.88 & 0.85 & 444 & -1.23 & 1.05  \\ 
        Genetics & 756 & -0.55 & 0.87 & 364 & -0.13 & 0.61 & 377 & -0.61 & 0.79  \\ 
        Physics 1 & 1167 & -0.35 & 0.61 & 637 & -0.24 & 0.61 & 893 & -0.64 & 0.79  \\ 
        Physics 2 & 764 & -0.30 & 0.63 & 527 & -0.18 & 0.62 & 623 & -0.35 & 0.73  \\ 
        Human Physiology & 901 & -0.32 & 0.74 & 407 & -0.14 & 0.63 & 799 & -0.56 & 0.81  \\ 
        Biochemistry & 523 & -0.96 & 0.89 & 399 & -0.13 & 0.66 & 564 & -0.67 & 1.00 \\   \bottomrule
    \end{tabular}}
\raggedright
\bigskip 
\small\textit{Note}. Mean and standard deviation (SD) of average grade anomalies (AGA), as well as number of students (N) for each course of interest.
\end{table}

\begin{table}[h]
\tbl{Average grade anomalies (AGAs), grades, and between-gender effect sizes for each course of interest in the four semesters before the COVID-19 pandemic. \label{AGA_grade_gender_precovid}}
{\begin{tabular}{l|ccccc|ccccc|cc} \toprule
 &  \multicolumn{5}{c}{Women} & \multicolumn{5}{c}{Men} & & \\ \cmidrule{2-6} \cmidrule{7-11}
 & &   \multicolumn{2}{c}{AGA} & \multicolumn{2}{c}{Grade} & & \multicolumn{2}{c}{AGA} & \multicolumn{2}{c}{Grade} & \multicolumn{2}{c}{Cohen's \textit{d}} \\
        Course & N & Mean & SD & Mean & SD & N & Mean & SD & Mean & SD & AGA & Grade \\ \midrule
        Calculus 1 & 385 & -0.79 & 1.48 & 2.58 & 1.11 & 436 & -0.92 & 1.90 & 2.34 & 1.30 & 0.07 & \textbf{0.20$^{**}$}   \\ 
        Biology 1 & 1077 & -0.70 & 1.17 & 2.86 & 1.02 & 571 & -0.55 & 0.92 & 2.88 & 1.00 & \textbf{-0.13$^{*}$} & -0.02  \\ 
        Biology 2 & 898 & -0.39 & 0.65 & 3.09 & 0.65 & 517 & -0.30 & 0.63 & 3.11 & 0.79 & \textbf{-0.13$^{*}$} & -0.03  \\ 
        Introductory Chemistry 1 & 1166 & -0.79 & 1.38 & 2.86 & 0.87 & 677 & -0.61 & 1.41 & 2.82 & 0.97 &\textbf{-0.13$^{**}$}  & 0.04  \\ 
        Introductory Chemistry 2  & 973 & -0.58 & 0.76 & 2.94 & 0.86 & 602 & -0.53 & 0.86 & 2.85 & 0.91 & -0.06 & 0.10  \\ 
        Organic Chemistry 1 & 1058 & -1.04 & 0.99 & 2.47 & 1.12 & 589 & -0.87 & 0.94 & 2.57 & 1.07 & \textbf{-0.18$^{***}$} & -0.10    \\ 
        Organic Chemistry 2  & 687 & -0.96 & 0.92 & 2.55 & 1.07 & 420 & -0.83 & 0.98 & 2.64 & 1.17 & \textbf{-0.13$^{*}$} & -0.08  \\ 
        Genetics & 449 & -0.47 & 0.76 & 3.02 & 0.94 & 307 & -0.67 & 1.00 & 2.75 & 1.18 &\textbf{0.23$^{**}$} & \textbf{0.26$^{***}$} \\ 
        Physics 1 & 731 & -0.41 & 0.60 & 3.06 & 0.76 & 435 & -0.26 & 0.63 & 3.13 & 0.84 & \textbf{-0.25$^{***}$} & -0.09  \\ 
        Physics 2 & 457 & -0.32 & 0.61 & 3.19 & 0.77 & 289 & -0.27 & 0.67 & 3.15 & 0.87 & -0.07 & 0.04  \\ 
        Human Physiology & 568 & -0.36 & 0.76 & 3.18 & 0.90 & 333 & -0.27 & 0.71 & 3.20 & 0.87 & -0.12 & -0.03  \\ 
        Biochemistry &  311 & -0.94 & 0.86 & 2.59 & 1.04 & 212 & -0.99 & 0.91 & 2.42 & 1.13 & 0.06 & 0.15 \\  \bottomrule
    \end{tabular}}
    \raggedright
\bigskip 
\small\textit{Note}. Cohen's \textit{d} is positive if women had higher grades or smaller AGAs than men in a course. A bold Cohen's \textit{d} signifies that a \textit{t}-test showed significant differences between men and women.\\
$^{*} = p < 0.05$, $^{**} = p < 0.01$, and $^{***} = p < 0.001$.
\end{table}

\begin{table}[h]
\tbl{Average grade anomalies (AGAs), grades, and between-gender effect sizes for each course of interest in the two semesters of remote instruction due to the COVID-19 pandemic. \label{AGA_grade_gender_covid}}
{\begin{tabular}{l|ccccc|ccccc|cc} \toprule
 &  \multicolumn{5}{c}{Women} & \multicolumn{5}{c}{Men} & & \\ \cmidrule{2-6} \cmidrule{7-11}
 & &   \multicolumn{2}{c}{AGA} & \multicolumn{2}{c}{Grade} & & \multicolumn{2}{c}{AGA} & \multicolumn{2}{c}{Grade} & \multicolumn{2}{c}{Cohen's \textit{d}} \\
        Course & N & Mean & SD & Mean & SD & N & Mean & SD & Mean & SD & AGA & Grade \\ \midrule
        Calculus 1 & 184 & -0.69 & 0.94 & 2.73 & 0.93 & 179 & -0.48 & 0.84 & 2.81 & 0.87 & \textbf{-0.24$^{*}$} & -0.08  \\ 
        Biology 1 & 305 & -0.49 & 0.74 & 3.05 & 0.81 & 166 & -0.30 & 0.70 & 3.10 & 0.88 & \textbf{-0.26$^{**}$} & -0.05  \\ 
        Biology 2 & 352 & -0.15 & 0.64 & 3.29 & 0.78 & 172 & -0.08 & 0.62 & 3.32 & 0.76 & -0.12 & -0.03  \\ 
        Introductory Chemistry 1 & 358 & -0.36 & 0.64 & 3.17 & 0.66 & 191 & -0.34 & 0.69 & 3.04 & 0.78 & -0.03 & \textbf{0.19$^{*}$}  \\ 
        Introductory Chemistry 2 & 352 & -0.70 & 0.69 & 2.81 & 0.82 & 203 & -0.44 & 0.65 & 2.81 & 0.87 & \textbf{-0.22$^{*}$}  & 0.00  \\ 
        Organic Chemistry 1 & 497 & -0.82 & 0.79 & 2.78 & 0.91 & 244 & -0.69 & 0.71 & 2.84 & 0.88 & \textbf{-0.17$^{*}$} & -0.07  \\
        Organic Chemistry 2 & 411 & -0.91 & 0.81 & 2.69 & 0.98 & 197 & -0.81 & 0.93 & 2.75 & 1.09 & -0.13 & -0.05  \\
        Genetics & 229 & -0.13 & 0.61 & 3.42 & 0.74 & 135 & -0.13 & 0.61 & 3.39 & 0.77 & 0.00 & 0.04 \\ 
        Physics 1 & 390 & -0.28 & 0.60 & 3.24 & 0.75 & 247 & -0.16 & 0.62 & 3.31 & 0.80 & \textbf{-0.20$^{*}$} & -0.09  \\ 
        Physics 2 & 330 & -0.23 & 0.67 & 3.36 & 0.73 & 197 & -0.09 & 0.53 & 3.47 & 0.71 & \textbf{-0.24$^{**}$} & -0.16   \\ 
        Human Physiology & 262 & -0.15 & 0.61 & 3.45 & 0.71 & 145 & -0.12 & 0.66 & 3.47 & 0.80 & -0.04 & -0.03  \\ 
        Biochemistry & 241 & -0.17 & 0.65 & 3.46 & 0.76 & 158 & -0.08 & 0.67 & 3.45 & 0.80 & -0.14 & 0.01 \\ \bottomrule
    \end{tabular}}
    \raggedright
\bigskip 
\small\textit{Note}. Cohen's \textit{d} is positive if women had higher grades or smaller AGAs than men in a course. A bold Cohen's \textit{d} signifies that a \textit{t}-test showed significant differences between men and women.\\
$^{*} = p < 0.05$, $^{**} = p < 0.01$, and $^{***} = p < 0.001$.
\end{table}

\begin{table}[h]
\tbl{Average grade anomalies (AGAs), grades, and between-gender effect sizes for each course of interest in the two semesters of in-person instruction after remote classes due to COVID-19. \label{AGA_grade_gender_postcovid}}
{\begin{tabular}{l|ccccc|ccccc|cc} \toprule
 &  \multicolumn{5}{c}{Women} & \multicolumn{5}{c}{Men} & & \\ \cmidrule{2-6} \cmidrule{7-11}
 & &   \multicolumn{2}{c}{AGA} & \multicolumn{2}{c}{Grade} & & \multicolumn{2}{c}{AGA} & \multicolumn{2}{c}{Grade} & \multicolumn{2}{c}{Cohen's \textit{d}} \\
        Course & N & Mean & SD & Mean & SD & N & Mean & SD & Mean & SD & AGA & Grade \\ \midrule
        Calculus 1  & 123 & -1.35 & 1.53 & 1.87 & 1.41 & 165 & -0.76 & 1.51 & 2.35 & 1.36 & \textbf{-0.39$^{**}$} & \textbf{-0.35$^{**}$}   \\ 
        Biology 1 & 139 & -0.88 & 1.15 & 2.30 & 1.19 & 73 & -0.37 & 0.87 & 2.66 & 0.98 & \textbf{-0.47$^{**}$} &\textbf{-0.32$^{*}$}  \\ 
        Biology 2 & 112 & -0.58 & 0.64 & 2.76 & 0.82 & 90 & -0.44 & 0.73 & 2.78 & 0.94 & -0.19 & -0.02   \\ 
        Introductory Chemistry 1& 133 & -0.83 & 1.12 & 2.46 & 1.06 & 122 & -0.62 & 0.88 & 2.48 & 0.97 & -0.21 & -0.02  \\ 
        Introductory Chemistry 2 & 171 & -1.12 & 0.92 & 2.21 & 1.08 & 134 & -0.82 & 0.83 & 2.38 & 1.10 & \textbf{-0.35$^{**}$} & -0.16  \\ 
        Organic Chemistry 1 & 331 & -1.28 & 0.95 & 2.12 & 1.12 & 191 & -0.73 & 0.97 & 2.61 & 1.14 &\textbf{-0.57$^{***}$} &\textbf{-0.44$^{***}$}  \\ 
        Organic Chemistry 2 & 269 & -1.43 & 1.00 & 2.01 & 1.19 & 175 & -0.92 & 1.07 & 2.47 & 1.27 & \textbf{-0.50$^{***}$} & \textbf{-0.38$^{***}$}   \\ 
        Genetics & 241 & -0.69 & 0.78 & 2.76 & 0.98 & 136 & -0.47 & 0.78 & 2.89 & 1.08 & \textbf{-0.28$^{**}$} & -0.13 \\ 
        Physics 1 & 601 & -0.73 & 0.78 & 2.73 & 0.94 & 292 & -0.47 & 0.77 & 2.89 & 0.98 & \textbf{-0.33$^{***}$}& \textbf{-0.16$^{*}$} \\ 
        Physics 2 Post-Remote & 374 & -0.37 & 0.73 & 3.16 & 0.91 & 249 & -0.31 & 0.74 & 3.17 & 0.99 & -0.09 & -0.01  \\ 
        Human Physiology & 516 & -0.60 & 0.80 & 2.95 & 0.98 & 283 & -0.50 & 0.82 & 3.06 & 1.00 & -0.11 & -0.10 \\ 
        Biochemistry & 371 & -0.69 & 1.00 & 2.82 & 1.22 & 193 & -0.65 & 1.02 & 2.79 & 1.29 & -0.03 & 0.03 \\ \bottomrule
    \end{tabular}}
    \raggedright
\bigskip 
\small\textit{Note}. Cohen's \textit{d} is positive if women had higher grades or smaller AGAs than men in a course. A bold Cohen's \textit{d} signifies that a \textit{t}-test showed significant differences between men and women.\\
$^{*} = p < 0.05$, $^{**} = p < 0.01$, and $^{***} = p < 0.001$.
\end{table}

\end{document}